%% file: main.tex
\input{style/acm_style}

\usepackage{amsmath,amsfonts}
\usepackage{hyperref}
\usepackage{cleveref}
\usepackage{multirow}
\usepackage{enumitem}
\usepackage{xcolor}
\setlist{nolistsep}
\usepackage{xspace}
\usepackage{booktabs}
\usepackage{xurl}
\usepackage[most]{tcolorbox}
\usepackage{fancyhdr}

\tcbset{
    colback=blue!5!white,
    colframe=blue!60!black,
    sharp corners,
    boxrule=0.6pt,
    boxsep=2pt, left=3pt, right=3pt, top=2pt, bottom=2pt
}
\newtcolorbox{fallacybox}[1][]{
    fonttitle=\bfseries, breakable, #1
}

\input{syssec-template}

\begin{document}
\input{authors/author-acm}

\title{Understanding the Privacy-Preserving Potential of \httptwo\ Against Webpage Fingerprinting}
\ifextver
\subtitle{\small
    Full version of the paper accepted at ACM CCS 2026; includes extended appendices. Please cite the conference version.
}
\fi

\input{sections/0_abstract}
\maketitle
\ifextver
    \pagestyle{fancy} 
    \fancyfoot[C]{\thepage} 
\fi
\input{sections/1_introduction}
\input{sections/2_threat_model}

\input{sections/4_method_experiments/a_intro_fingerprinting}

\input{sections/4_method_experiments/b1_intro_defenses_benchmarks}
\input{sections/4_method_experiments/b2_intro_defenses}

\input{sections/4_method_experiments/c_ref_client_defenses}
\input{sections/4_method_experiments/d_ref_server_defenses}
\input{sections/4_method_experiments/e_http2_opportunities_analysis}
\input{sections/6_conclusion}
\bibliographystyle{ACM-Reference-Format}

\bibliography{references}

\input{sections/7_appendix_A_openscience}
\input{sections/7_appendix_B_ethics}

\ifextver
    \input{sections/7_appendix_D_implementation}

    \input{sections/7_appendix_E_calibration}

\else
    \input{sections/7_appendix_C_extras}
\fi
\end{document}

%% file: style/acm_style.tex
\newif\ifextver

\extvertrue     

\ifextver
    \documentclass[sigconf, nonacm]{acmart}
\else
    \documentclass[sigconf]{acmart}
\fi

\ifextver
\else
    \acmConference[CCS '26]
        {2026 ACM SIGSAC Conference on Computer and Communications Security}
        {November 15--19, 2026}
        {The Hague, The Netherlands}
    \copyrightyear{2026}
    \acmYear{2026}
    \acmDOI{XXXXXXX.XXXXXXX}
    \setcopyright{acmlicensed}
    \acmISBN{978-1-4503-XXXX-X/2018/06}
\fi

%% file: syssec-template.tex
\usepackage{xcolor}
\usepackage[normalem]{ulem}
\usepackage{pifont}
\usepackage{soul}

\newcommand{\cmark}{\ding{52}}
\newcommand{\xmark}{\ding{53}}

\newcommand{\http}{\texttt{HTTP}\xspace}
\newcommand{\https}{\texttt{HTTPS}\xspace}
\newcommand{\tls}{\texttt{TLS}\xspace}
\newcommand{\ip}{\texttt{IP}\xspace}
\newcommand{\dns}{\texttt{DNS}\xspace}
\newcommand{\udp}{\texttt{UDP}\xspace}
\newcommand{\tcp}{\texttt{TCP}\xspace}
\newcommand{\quic}{\texttt{QUIC}\xspace}
\newcommand{\tor}{\texttt{Tor}\xspace}
\newcommand{\vpn}{\texttt{VPN}\xspace}
\newcommand{\httpone}{\texttt{HTTP/1.1}\xspace}
\newcommand{\httptwo}{\texttt{HTTP/2}\xspace}
\newcommand{\httpthree}{\texttt{HTTP/3}\xspace}

%% file: authors/author-acm.tex
\author{Bogdan Cebere}
\email{bogdan.cebere@cispa.de}
\affiliation{%
   \institution{CISPA Helmholtz Center for Information Security}
   \city{Dortmund}
   \country{Germany}
}

\author{Prateek Kumar}
\email{prateek21081@iiitd.ac.in}
\affiliation{%
   \institution{CISPA Helmholtz Center for Information Security}
   \city{Saarbrücken}
   \country{Germany}
}

\author{Sylvain Chatel}
\email{sylvain.chatel@cispa.de}
\affiliation{%
   \institution{CISPA Helmholtz Center for Information Security}
   \city{Saarbrücken}
   \country{Germany}
}

\author{Wouter Lueks}
\email{lueks@cispa.de}
\affiliation{%
   \institution{CISPA Helmholtz Center for Information Security}
   \city{Saarbrücken}
   \country{Germany}
}

\author{Christian Rossow}
\email{rossow@cispa.de}
\affiliation{%
   \institution{CISPA Helmholtz Center for Information Security}
   \city{Dortmund}
   \country{Germany}
}

%% file: sections/0_abstract.tex
\begin{abstract}

Website fingerprinting (WF) attacks can infer which webpage a user visits from encrypted \https traffic alone, compromising privacy even without decryption.
WF defenses commonly shape traffic through noise, padding, delays, or flow splitting --- yet they are most often studied from the perspective of encapsulating protocols like \tor or \vpn rather than at the application layer (\http). 

In this work, we focus on application-layer defenses enabled by the most widely deployed version of \http\ --- \httptwo. We demonstrate how known defenses can be emulated through \httptwo features at the client side (HTTPOS, LLaMA, FRONT, Tamaraw) and the server side (ALPaCA, Tamaraw). We further show that \httptwo features --- such as proactive resource suggestion, multiplexing, and flow control --- offer untapped potential for lightweight yet effective defenses deployable at both endpoints.

We evaluate these defenses using a unified blueprint that calibrates defense parameters per dataset, then combines practical attacks, information-theoretic leakage estimates, and overhead measurements.
For each defense, this framework identifies the strongest hyperparameter-tuned fingerprinting model and estimates the residual uncertainty induced by the defense using two information-theoretic leakage estimators --- all while accounting for the defense's privacy--overhead trade-offs.

\end{abstract}

\ccsdesc[500]{Security and privacy~Web protocol security}
\ccsdesc[300]{Computing methodologies~Machine learning}

\keywords{Website Fingerprinting, Machine Learning, HTTP/2} 

%% file: sections/1_introduction.tex
\section{Introduction}
\label{section:introduction}

Website fingerprinting is a traffic analysis technique that enables adversaries to infer users’ online activities --- such as political or social behavior --- and to facilitate adversarial behavior, including censorship and targeted surveillance. Website fingerprinting remains feasible despite the widespread deployment of \http\ over \tls\ (i.e., \https), since \tls\ encrypts only the application payload, leaving network- and transport-layer headers and traffic characteristics (packet sizes, timings, and ordering) observable to on-path adversaries. Website fingerprinting leverages this residual metadata to reconstruct browsing patterns and compromise user privacy.

In response, fingerprinting defenses aim to make the \ip/\tcp/\udp metadata uninformative. Popular obfuscation techniques rely on the \tor and \vpn protocols, which encapsulate traffic and obscure its content from on-path observers.
However, most web traffic does not use \tor or \vpn, leaving \http packet metadata directly exposed to on-path adversaries such as ISPs or network operators. 
For this reason, we focus on fingerprinting scenarios that can be defended by clients or servers at the application-layer (\http).

Application-layer fingerprinting defenses generally pursue one of two strategies: (1) uniformity, wherein the defense transforms webpage metadata into fixed packet lengths and rates; or (2) unpredictability, wherein the defense ensures that metadata differs across subsequent page loads.
To follow these strategies, fingerprinting defenses commonly rely on four obfuscation functionalities: (1) noise traffic insertion, (2) packet padding, (3) packet delay, and (4) packet splitting \cite{DBLP:conf/uss/SmithDMP22}. 
These mechanisms do not hide \ip-layer metadata or the destination domain and therefore complement (not replace) \tor or \vpn solutions when network-layer anonymity is required.
Instead, they target subpage fingerprinting over ordinary \https by perturbing transport-visible features such as packet sizes, timing, and ordering, thereby reducing leakage about the specific content accessed within a known domain.
This distinction is relevant when the domain itself is not sensitive, but the visited subpage is, such as a particular product or medical page.

The privacy potential of modern \http-layer features like multiplexing, reprioritization, and flow control has not been fully explored yet. Prior work has shown that established fingerprinting defenses can be emulated in \httpthree \cite{DBLP:conf/uss/SmithDMP22}, but no equivalent study exists for \httptwo. 
This gap is practically important: as of 2026, \httptwo remains the predominant deployed \http version, accounting for approximately $60\%$ of browser traffic, compared with roughly $31\%$ for \httpthree \cite{cloudflareExaminingHTTP3, cloudflareRadarAdoption2026}. Thus, \httptwo is not merely a legacy target, but the protocol on which the majority of today's Web traffic still depends.
Moreover, the \httptwo defenses are more challenging to design, as \tcp's flow control is typically handled by the OS kernel, whereas \quic (\httpthree) implementations commonly run in userspace, allowing applications more direct control over flow-control–driven traffic shaping.
Beyond emulating existing defenses, \httptwo also opens the door to new strategies --- such as leveraging proactive resource hints or stream multiplexing --- but this potential likewise remains largely unexplored.

Before such defenses can be meaningfully assessed, we must first confront a fundamental challenge: \emph{how to measure whether they truly provide privacy}.
Evaluating defenses is inherently harder than evaluating attacks. While attacks can often succeed with a single model and dataset, defenses must be tested against multiple adversaries to establish credible security guarantees. Yet the common practice of reporting reduced fingerprinting accuracy under defense can be misleading, as degradation may also result from preprocessing errors, poor hyperparameter tuning, or near-misses by attackers.
Prior work has proposed complementary techniques, including privacy-overhead calibration curves \cite{DBLP:conf/ccs/CaiNWJG14} and model-agnostic estimators of information leakage \cite{DBLP:conf/ccs/LiGH18,DBLP:journals/popets/VeichtRB23,DBLP:journals/popets/Cherubin17}.
Building on these ideas, we propose a unified framework that calibrates each defense per dataset to select a practical privacy--overhead operating point, then stress-tests the selected configuration against hyperparameter-tuned attackers and leakage estimators.

\noindent\textbf{Contributions.} This paper explores how \httptwo can be harnessed for website fingerprinting defenses. Our contributions are threefold:
\begin{enumerate}[left=0pt,noitemsep]
    \item \textbf{Evaluation framework}. We introduce a benchmarking blueprint for assessing the quality of fingerprinting defenses that unifies practical attacks, information-theoretic leakage estimates, and overhead measurement (\Cref{section:wf_defenses_benchmarks}). For each defense, the framework calibrates its operating parameters per dataset, evaluates its privacy--overhead trade-off, identifies the strongest hyperparameter-tuned attack, and measures the residual leakage that remains.
    \item \textbf{Defense emulation.} We demonstrate how established defenses --- both client-side \cite{DBLP:conf/uss/GongW20,DBLP:conf/ccs/CaiNWJG14,DBLP:conf/ndss/LuoZCLCP11,DBLP:journals/popets/CherubinHJ17} and server-side \cite{DBLP:journals/popets/CherubinHJ17,DBLP:conf/ccs/CaiNWJG14} --- can be reproduced using \httptwo primitives, and evaluate their effectiveness on real-world case studies using our framework (\Cref{section:wf_defenses_emulation}). 
    The results show substantial but dataset-dependent privacy gains: even the strongest attackers cannot narrow the set of plausible webpages to fewer than $26$ candidates on four of five datasets with client-side defenses, or fewer than $63$ candidates on any dataset with appropriately placed server-side defenses, whether at the $1^\text{st}$-party server, CDN, or across all involved servers.
    \item \textbf{New opportunities.} We explore untapped \httptwo features for privacy, highlighting novel defense strategies such as client-side multiplexing and server-side proactive resource suggestion (\Cref{section:http2_potential}). 
    These client-side techniques provide more consistent protection across datasets, keeping the attacker's estimated candidate set above seven webpages in \emph{all} five cases, while reducing latency overhead by $55$--$98\%$ and downstream overhead by $20$--$66\%$.
    On the server side, \httptwo enables the $1^\text{st}$-party server to deploy defenses without controlling all content-delivery servers, keeping the plausible candidate set above $13$ webpages across all datasets while reducing bandwidth overhead by $54$--$80\%$ on most datasets.    
    
\end{enumerate}

%% file: sections/2_threat_model.tex

\section{Threat Model and Attacker Goals}
\label{subsection:threat_model}
We consider a passive on-path adversary with access to the \tcp layer, who attempts to determine the visited webpage from an encrypted \httptwo traffic trace by analyzing traffic characteristics such as packet lengths, inter-arrival times, and ordering. We assume the adversary can observe individual \tcp connections during a page load (one per subdomain involved in the page load), grouping them into a single \emph{trace} while keeping each connection identifiable.

We assume the adversary cannot alter the communication -- i.e., they do not actively intercept the communication using Machine-in-the-Middle attacks. This scenario is common (e.g., Internet or VPN providers observing communication) and more stealthy --- no need to install custom certificates on the client. However, we also assume a defense-aware adversary who knows which defense mechanism is deployed and can collect labeled defended traces for training. 

We focus on \emph{subpage fingerprinting}: identifying which specific subpage of a known domain a user visits --- for example, the exact Amazon product page, the particular BBC article, or the specific Reddit thread. 
In this setting, all contacted subdomains during webpage load are already observable via \ip addresses, \dns resolves, and \tls handshakes.
In principle, the sequence of subpage-specific \dns lookups and \tls SNI values for third-party services (e.g., CDNs) could additionally leak subpage identity, but such leakage is 
orthogonally mitigated by DNS-over-HTTPS (DoH)~\cite{ietf8484Queries} or Encrypted Client Hello (ECH)~\cite{ietf9849Encrypted}. 
Specific to our empirical studies, assuming legacy clients without these features, \dns contact ordering alone yields near-random subpage classification on four of five datasets (F1\,$\leq 0.10$), as we show empirically in ~\Cref{subsubsection:benchmarks_dns_sni}. 
The \http layer, therefore, remains the primary actionable source of leakage in our threat model and the only one that application-layer \httptwo defenses can address.
Note that DoH and ECH can mitigate DNS/SNI leakage, but do not mitigate subpage fingerprinting as studied here.

%% file: sections/4_method_experiments/a_intro_fingerprinting.tex
\newcommand{\numdatasets}{five\xspace}
\newcommand{\numpages}{100\xspace}
\newcommand{\numrepeats}{500\xspace}

\section{Website Fingerprinting}
\label{section:baseline_website_fingerprinting}

To motivate the need for \httptwo privacy techniques, we first introduce established website fingerprinting techniques (\Cref{subsection:fingerprinting_techniques}), then apply them on real-world sensitive case studies (\Cref{subsection:realworld_dataset_baselines}).


\subsection{Fingerprinting Techniques} 
\label{subsection:fingerprinting_techniques}

In website fingerprinting, an observer attempts to extract sensitive information from encrypted communication by using network-layer information (e.g., IP addresses), transport-layer metadata (e.g., TCP segment sizes, connection counts), and derived traffic statistics (e.g., throughput, inter-packet timing, burst patterns) \cite{cheng1998traffic,sun2002statistical,DBLP:conf/uss/HayesD16,hintz2002fingerprinting,DBLP:conf/raid/CebereR24,DBLP:conf/ccs/CadenaMHPRFEWP20,DBLP:conf/ndss/PanchenkoLPEZHW16, deng2025countmamba, DBLP:conf/uss/0001JG0Z023}. This metadata is used to find the most similar page in an already-collected training set using Machine-Learning methods. Previous website fingerprinting works explored the potential of Gaussian Distributions \cite{DBLP:conf/pet/MillerHJT14}, linear models \cite{DBLP:journals/ccr/CoullD14}, random forests \cite{DBLP:conf/iscc/LiuCLX18}, clustering methods \cite{DBLP:conf/icc/ShenLCZZ19,DBLP:conf/uss/HayesD16}, and neural networks (NN) \cite{DBLP:conf/infocom/LiuHXCL19,DBLP:journals/tifs/ShenLZDH21,van_ede_flowprint_2020,DBLP:conf/lcn/DahanayakaJS20,DBLP:conf/uss/CuiGX0C021,shen_efficient_2021,DBLP:conf/www/LinXGLSY22,DBLP:conf/uss/XieCDX00SZ23,DBLP:journals/popets/BhatLKD19,DBLP:conf/ccs/Deng0024,DBLP:conf/ccs/SirinamIJW18}.
Most prior works focus on domain fingerprinting (using as labels the hosting domain, e.g., google.com or fb.com) under \tor encapsulation; some also explore the subpage-fingerprinting problem \cite{zhao2024towards,DBLP:journals/tifs/ShenLZDH21,zhang_deep_2019}.  In contrast to these works on subpage WF, our work focuses on subpage-WF in \httptwo (previously discussed only in \cite{zhang_deep_2019} on multiplexing effect).  

\subsubsection{\textbf{Fingerprinting Models}} 
\label{sub:fingerprinting_methods}

For the attacker effectiveness benchmarks, we evaluate each dataset against five fingerprinting models:
(1) \emph{k-FP}~\cite{DBLP:conf/uss/HayesD16}, which combines $k$-nearest neighbor clustering with a Random Forest;
(2) \emph{Deep Fingerprinting (DF)}~\cite{DBLP:conf/ccs/SirinamIJW18}, a CNN-based architecture that operates directly on raw traces;
(3) \emph{VarCNN}~\cite{DBLP:journals/popets/BhatLKD19}, a ResNet-style architecture designed to capture multi-scale burst patterns; 
(4) \emph{Holmes}~\cite{DBLP:conf/ccs/Deng0024}, which combines spatial--temporal pattern analysis with contrastive learning to identify websites before the page fully loads;
(5) \emph{Robust-Fingerprinting (RobustFP-CNN)}~\cite{DBLP:conf/uss/0001JG0Z023}, for which we reuse only the CNN classifier architecture, applying 2D convolutions over local packet windows to jointly capture packet-size and timing patterns in defended traces. We do not use RF's time-aggregated representation, which improves robustness to padding but would discard the informative variable packet lengths present in our \httptwo traces; consequently, RobustFP-CNN does not evaluate the full RF approach.

\noindent\emph{Note}: All WF models above were originally developed for \tor traffic; in contrast, in our benchmarks, the packet lengths are also informative (compared to constant-size \tor cells), and omitting them would artificially cripple the models. \emph{Therefore, the benchmarks are executed on different dataset types and features than those in the related work}.
\ifextver
Implementation details in \Cref{appendix:security_estimators_extra}.
\fi

\subsubsection{\textbf{Dataset Creation}} 
\label{sub:dataset_creation}

Throughout the empirical studies, we evaluate \httptwo defenses from both client and server perspectives, requiring control over both endpoints while replaying the same real webpage resources.
\ifextver
For each webpage, we first collect and cache browser requests and responses, including headers and body content (\Cref{appendix:browser_crawlers}), then replay them through a custom Python client-server with defenses enabled or disabled, (\Cref{appendix:client_server_details}) and capture the resulting PCAP traces (\Cref{appendix:pcap_captures}).
\else
For each webpage, we first collect and cache browser requests and responses, including headers and body content, then replay them through a custom Python client-server with defenses enabled or disabled, and capture the resulting PCAP traces.
\fi
This replay setup is necessary to evaluate server-side defenses under identical content, enable same-resource comparisons, and isolate \httptwo-layer effects from content and network-path variability.
The resulting measurements therefore characterize defense-induced changes rather than live-Web variability: browser/CDN scheduling and resource dependencies are not reproduced, which may affect latency and intra-page variance but does not undermine the controlled comparison targeted by our study.
Each PCAP trace is represented as a collection of connections, where each connection is a sequence of packet lengths, inter-arrival timings, and directions.

For every webpage and defense scenario, we generate $500$ defended PCAP traces, independently resampling all randomized defense choices for each replay. 
Under five-fold cross-validation, each attacker is therefore trained on approximately $400$ defended realizations per webpage. 
This provides hundreds of directly measured realizations of the defense-induced distribution for every class, so additional synthetic augmentation is not needed to compensate for limited training diversity. Moreover, artificially perturbing packet sizes or timings could introduce traces that deviate from the traffic distribution produced by the evaluated defense.

The traces are converted into two distinct evaluation datasets: a 2D representation (for k-FP), and a 3D representation (for DF, VarCNN, Holmes, and RobustFP-CNN models).
k-FP operates on handcrafted per-connection statistics: packet counts and unique sizes per direction; top-$N$ burst size statistics; cumulative incoming/outgoing byte totals interpolated to a fixed-length vector; and inter-arrival timing statistics. Each trace is first converted to an array of statistics, and the final dataset is a matrix of shape $(\text{n\_traces},\ \text{LIM}_{\text{conns}} * \text{n\_features})$. In our datasets, the bulk of the leakage is concentrated in at most three connections, so we set $\text{LIM}_{\text{conns}} = 3$.
The neural network models operate on a raw 3D time-series tensor with two channels: signed packet lengths (positive for upload, negative for download) and inter-arrival times. This differs from the original NN model designs~\cite{DBLP:conf/ccs/SirinamIJW18, DBLP:journals/popets/BhatLKD19, DBLP:conf/ccs/Deng0024}, which were built for \tor traffic where cell sizes are fixed and only direction, counts, and inter-arrival times are informative. We therefore include packet lengths as a dedicated channel in the tensor representation, acknowledging that this is not a direct architectural comparison but rather an adaptation to the richer information available at the \tcp layer. 
Within each trace, connection blocks are concatenated in arrival order, ordered by the timestamp of each connection's first packet.
Inter-arrival times are recorded relative to the start of each connection, so the timestamp resets to zero at every connection boundary; this acts as an implicit connection-boundary signal that the model can learn from.
Following prior work~\cite{DBLP:journals/popets/BhatLKD19, DBLP:journals/popets/RahmanSMG020, rimmer_automated_2018, DBLP:conf/ccs/SirinamIJW18}, we treat the maximum trace length $L$ as a hyperparameter and set $L = 5000$ packets, which does not 
truncate any trace in our benchmarks.
Each model outputs a probability distribution over the $N$ webpages in the training set.

\subsubsection{\textbf{Fingerprinting Benchmarks}} 
\label{sub:fingerprinting_benchmarks}
\newcommand{\TP}{\mathrm{TP}}
\newcommand{\FP}{\mathrm{FP}}
\newcommand{\FN}{\mathrm{FN}}

The primary objective of the fingerprinting benchmarks is \emph{to assess the effectiveness of the attacker}, represented by the fingerprinting model.
Given $C$ website classes, we evaluate each model using cross-validation $F1_c=2\TP_c/(2\TP_c+\FP_c+\FN_c)$ and report $\mathrm{Macro\text{-}F1}=C^{-1}\sum_{c=1}^{C}F1_c$,
where $\TP_c$, $\FP_c$, and $\FN_c$ denote the true positives, false positives, and false negatives for website $c$, respectively.
For each benchmark, we report the mean score and the 95\% confidence interval (mean $\pm \text{CI}_{95\%}$) across all folds.
As a rule of thumb, when using $C=100$ classes, the macro-F1 value is $\sim 0.01$ for random guesses, $\sim 0.1 - 0.5$ for weak attackers, and $0.8+$ for strong attackers.

\subsubsection{\textbf{Datasets Types}}
We investigate website fingerprinting against users who browse without additional encapsulation layers such as \tor or \vpn. 
Most website fingerprinting research focuses on \emph{domain-level} inference --- determining which domain a user visits --- using datasets such as Tranco or Alexa \cite{pochat2018tranco,DBLP:conf/pet/MillerHJT14,DBLP:conf/uss/CherubinJT22,DBLP:conf/webist/MartinoRQL18,DBLP:conf/ndss/PanchenkoLPEZHW16}. To avoid trivial leakage via \tls certificates or \ip addresses, these studies typically assume clients use anonymity systems like \tor or \vpn, which conceal both endpoints and traffic features.
In contrast, \emph{subpage fingerprinting} seeks to identify which specific subpage of a known domain a user visits (e.g., a product page on Amazon). This setting is more fine-grained, as subpages usually share similar \dns, \tls, and \ip traits, leaving most differences at the \http layer.

These architectural differences also influence how defenses behave. In \tor or \vpn, a passive observer sees a single long-lived tunnel that encapsulates all \tls connections. Thus, a defense applied to any one \tls connection modifies the metadata of the entire tunnel --- for example, protected traffic from one connection may overlap with the \tls handshake of another. 
In contrast, with plain \https (without \tor or \vpn), each connection is exposed individually --- therefore, it must be protected independently. Since our work targets plain \https traffic, we focus on subpage-fingerprinting case studies to evaluate \emph{application-layer defenses}.


\subsection{Fingerprinting Case Studies}
\label{subsection:realworld_dataset_baselines}

We now discuss \numdatasets fingerprinting case studies and their privacy risks.
These results will lay the foundation for the defensive benchmarks in the following sections.

\begin{table}[t]
\caption{Real-world datasets and their average number of unique servers, requests, the image count, and their sizes. }
\label{tab:realworld_dataset_summaries}
\centering
\resizebox{\columnwidth}{!}{
    \begin{tabular}{crrrr}
    \toprule
    \textbf{Dataset} & \textbf{\begin{tabular}[c]{@{}c@{}}\# Servers Used\\ per Page\end{tabular}} & \textbf{\begin{tabular}[c]{@{}c@{}}\# Avg. Req.\\ per Conn.\end{tabular}} & \textbf{\begin{tabular}[c]{@{}c@{}}\# Avg. Images \\ per Page\end{tabular}} & \textbf{\begin{tabular}[c]{@{}c@{}}Avg. Image \\ Size (KB)\end{tabular}} \\ \midrule
    Amazon           & $4.14 \pm 0.1$                                                           & $19.3 \pm 1.2$                                                              & $29.22 \pm 1.5$                                                            & $101.8 \pm 17.2$                                                         \\
    BBC              & $8.03 \pm 0.1$                                                           & $6.97 \pm 0.3$                                                              & $12.8 \pm 16.3$                                                            & $197.04 \pm 173.4$                                                       \\
    Reddit           & $8.21 \pm 0.1$                                                           & $6.87 \pm 0.2$                                                              & $10.66 \pm 0.2$                                                            & $84.71 \pm 20.5$                                                         \\
    Udemy            & $2.03 \pm 0.1$                                                           & $5.72 \pm 0.1$                                                              & $8.75 \pm 1.3$                                                             & $16.38 \pm 4.4$                                                          \\
    Wiki             & $3.83 \pm 0.1$                                                           & $8.14 \pm 0.3$                                                              & $15.42 \pm 0.7$                                                            & $32.16 \pm 1.4$                                                          \\   \bottomrule
    \end{tabular}
}
\end{table}

\subsubsection{\textbf{Datasets}}
\label{subsubsection:baseline_selected_datasets}

We collect $100$ subpages from \numdatasets public websites --- Amazon, BBC, Reddit, Udemy, and Wikipedia --- as balanced datasets with $500$ samples per page --- using the steps detailed in \Cref{sub:dataset_creation}. 
These five websites are case studies rather than a representative sample of the Internet, chosen to capture variation in sensitive connections, requests per connection, and image counts.
This diversity is central to our analysis: both the calibrated defense parameters and the strongest hyperparameter-tuned attackers vary across datasets (\Cref{subsection:baseline_fingerprinting_defenses_clients}), showing that defense effectiveness is structure-dependent and that no single configuration is universally representative.
We therefore evaluate multiple websites to capture dataset-dependent defense behavior rather than draw general conclusions from a single dataset.
To assess the effectiveness of fingerprinting defenses, we focus on \emph{closed-world scenarios}, where the attacker has full knowledge of all target pages. This setting is intentionally chosen because it represents the most challenging case to defend, and an \emph{upper bound on attacker accuracy}.

\Cref{tab:realworld_dataset_summaries} summarizes some of the per-webpage statistics of each dataset: the average number of servers contacted per webpage, the average request count per connection, and the average image size observed in each webpage for each dataset. Per page load, each connection is a sequence of requests (e.g., images, script downloads) and can create a fingerprinting side-channel (alone or combined with other connections).
Due to larger resource-specific sizes, images are a primary source for fingerprinting, as previous studies also analyzed \cite{DBLP:conf/raid/CebereR24,DBLP:conf/www/WangZBKD21}, and thus we report their statistics separately.

\subsubsection{\textbf{Fingerprinting on the Undefended Datasets}}
\label{subsubsection:baseline_results}

\begin{figure}[t!]
        \centering
        \centerline{\includegraphics[width=\linewidth]{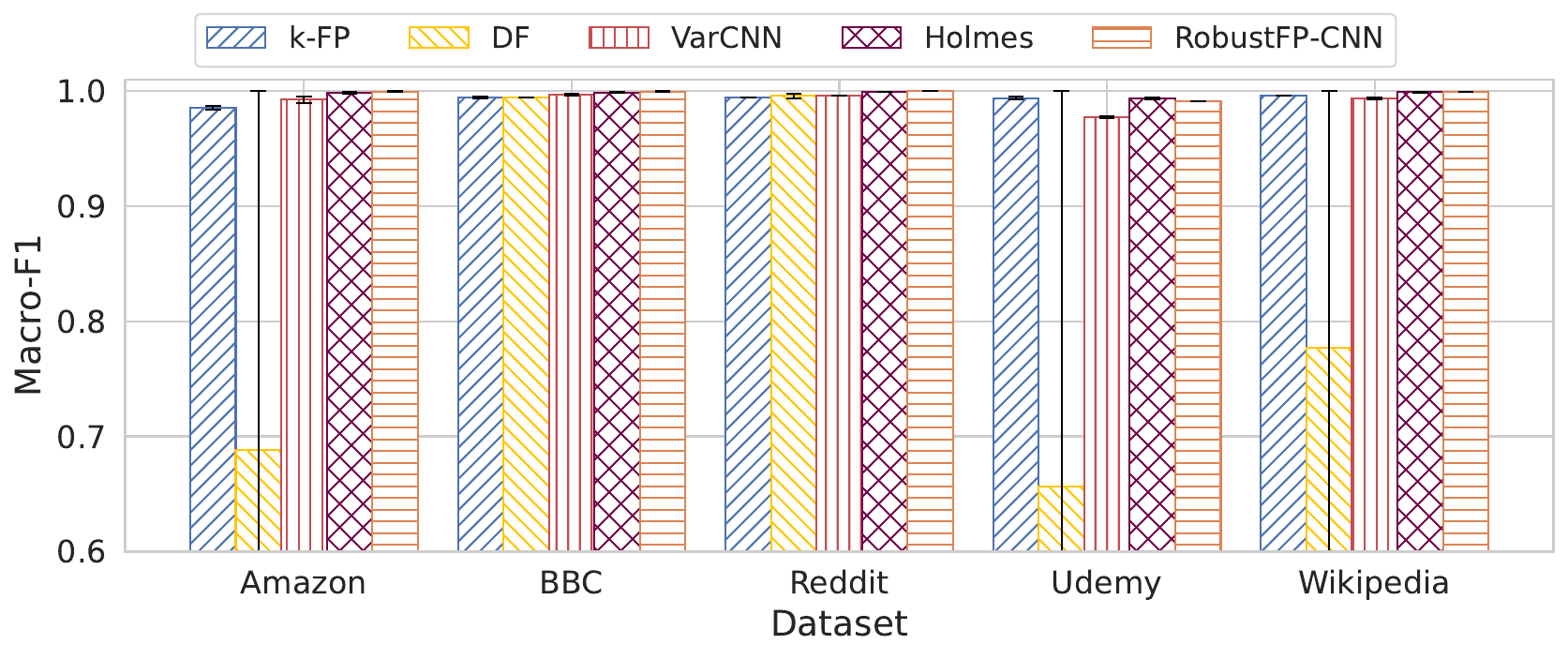}}
        \caption{Fingerprinting performance (F1 score) on the undefended real-world datasets. }
        \Description{Undefended baseline model performance}
        \label{fig:realworld_f1_score}
\end{figure}

\Cref{fig:realworld_f1_score} summarizes the fingerprinting performance of each model across the five datasets. 
We observe that, for every dataset, multiple hypertuned fingerprinting architectures achieve a macro-F1 score $\geq 0.9$, indicating high webpage-identification accuracy. 
In particular, strong performance is achieved by both feature-engineered approaches such as k-FP, which rely on global traffic statistics, and deep-learning approaches that learn representations directly from packet sequences. 
Overall, these results show that an effective defense must mitigate both feature-engineered and deep-learning-based fingerprinting attacks.

Intuitively, leakage sources group into: \emph{packet counts}, \emph{timing/IAT} (inter-arrival times), \emph{bursts} (how data is grouped, e.g., initial HTML followed by resources), and \emph{CUMUL} (the overall growth of the transfer, such as how upload or download bytes accumulate over time). 
Manual inspection shows per-site fingerprints resulted from:
\begin{itemize}[left=0pt,noitemsep]
    \item Amazon: number, size, and IATs of image requests to third-party CDNs (\texttt{media-amazon}, \texttt{ssl-images-amazon}).
    \item BBC: IAT, burst, and CUMUL across $\geq 2$ connections --- first-party plus a CDN (\texttt{static.files.bbci}).
    \item Reddit: burst and CUMUL on first-party and a CDN (\texttt{redditmedia}).
    \item Udemy: burst and CUMUL, almost entirely on the first-party connection (\texttt{www.udemy.com}).
    \item Wikipedia: first-party (\texttt{wikipedia.org}) dominates --- packet counts, bursts, and CUMUL from image downloads.
\end{itemize}

\subsubsection{\textbf{Scope of Leakage: \http vs. Sequence of Domains (\dns or SNI) }}
\label{subsubsection:benchmarks_dns_sni}
\httptwo is not, in itself, sufficient to protect the subpage's identity. For legacy clients --- without DoH or ECH support --- the target webpage may also be visible through other channels, such as the sequence of plaintext \dns queries or \tls SNI within the webpage load.
To assess whether \dns contact ordering alone enables fingerprinting, we encoded the sequence of contacted domain names during webpage load as features and trained a k-FP classifier on the dataset. The resulting macro-F1 scores are: Amazon $0.10$, BBC $0.07$, Reddit $0.74$, Udemy $0.01$, Wikipedia $0.03$. 
\dns contact ordering is therefore largely uninformative for subpage identity across four of five datasets. Reddit is the only outlier, likely due to its large number of \emph{unique} servers per subpage ($8.21 \pm 0.1$, Table~\ref{tab:realworld_dataset_summaries}), which causes subpage-specific TLS handshakes. BBC, despite a similarly large server count ($8.03 \pm 0.1$), contacts largely the same servers across subpages, leaving \dns uninformative.
Since \httptwo application-layer defenses cannot address \dns or \tls leakage --- that requires orthogonal mechanisms such as DoH or ECH --- we treat Reddit's \dns leakage as a known limitation and focus our defenses on the \http packet-level side channel, which dominates in four of five datasets.

\subsubsection*{\textbf{Takeaways}}
We presented \numdatasets\ datasets that exhibit diverse fingerprinting leakage characteristics: (i) some exhibit at least one dominant leaking subdomain/connection (Amazon, Udemy, Wikipedia); (ii) others involve multiple connections contributing to page identification (Reddit, BBC); (iii) certain datasets contain informative packet counts (Amazon, Wikipedia); 
(iv) some datasets leak information through processing duration (Amazon, BBC); 
and (v) all exhibit leakage through CUMUL and burst statistics. 
Any successful fingerprinting defense must address all of these side channels.

%% file: sections/4_method_experiments/b1_intro_defenses_benchmarks.tex
\newcommand{\benchkey}{BBQ\xspace}
\section{Benchmarking WF Defenses}
\label{section:wf_defenses_benchmarks}

Defense benchmarks aim to quantify how closely any attacker could infer the correct website.
However, relying solely on fingerprinting accuracy (\Cref{sub:fingerprinting_benchmarks}) can be misleading: a defense may only defeat specific models or provide privacy at the cost of impractically high overhead. Moreover, implementation errors such as faulty preprocessing or incorrect feature scaling can artificially lower an attacker’s performance, creating a false sense of security.

To address these pitfalls, we propose a novel Blueprint for Benchmarks and Quality Assurance (\benchkey) of fingerprinting defenses. 
The framework calibrates each defense and evaluates it using five practical attackers (k-FP, DF, VarCNN, RobustFP-CNN, Holmes), information-theoretic leakage estimates from two security estimators (WeFDE \cite{DBLP:conf/ccs/LiGH18} and DeepSE-WF \cite{DBLP:journals/popets/VeichtRB23}), and the privacy–overhead trade-off. We provide takeaways of our benchmarking blueprint at the end of \Cref{subsection:takeawayBlueprint}.

\noindent\fbox{\textbf{\benchkey0}} \textbf{Defense calibration}: \emph{Which defense configuration is recommended for this website and the underlying protocol (here: \httptwo)?}

\noindent Defense parameters inherited from prior protocols may not transfer directly to \httptwo, and multiple parameters may jointly affect privacy and overhead. We therefore define and evaluate several defense configurations at increasing intensity levels. Cai et al.~\cite{DBLP:conf/ccs/CaiNWJG14} similarly evaluated privacy--overhead trade-offs across defense parameter settings; however, their defenses were applied offline to packet sequences, whereas we implement each configuration in \httptwo and generate defended traces by replaying real webpages.

Using a calibration set containing 100 traces per webpage, we characterize each configuration by the maximum Macro-F1 across the five practical attackers in \benchkey1, together with bandwidth and latency overheads measured as in
\benchkey4. 
We refer to this maximum as the \emph{calibration Macro-F1}.
For each dataset, we then select a practical operating point from the observed privacy--overhead curve, favoring configurations that provide substantial reductions in calibration Macro-F1 while avoiding higher-intensity settings whose additional overhead yields limited privacy improvement.
The selected defense parameters are then fixed and evaluated under \benchkey1--4 on a separately generated dataset containing 500 traces per webpage.
\ifextver
\else
Additional implementation details and complete calibration sweeps are provided in the full version~\cite{XXX}.
\fi

\noindent\fbox{\textbf{\benchkey1}} \textbf{Practical Attacker Effectiveness:} \emph{How accurately does an attacker perform against the defense? }

\noindent For each defense, we measure and report the maximum Macro-F1 score (\Cref{sub:fingerprinting_benchmarks}) of the fingerprinting models introduced in \Cref{sub:fingerprinting_methods} --- the ``baseline resilience" of each defense. 
Each attacker is hyperparameter-tuned independently for the corresponding dataset--defense pair, retaining its default configuration when tuning does not improve validation Macro-F1.
Intuitively, the macro-F1 value is $\sim 0.1 - 0.6$ for strong defenses (against the evaluated attacker), and $0.8+$ for a weak defense.

\noindent\fbox{\textbf{\benchkey2}} \textbf{Practical Attacker Predictive Proximity:} \emph{How close are the attackers to the correct answer?}

\noindent The F1 scores hide if a model ``almost'' predicted a perfect match. 
Using the same tuned attacker configurations as in \benchkey1, we therefore also report Top-$k$ accuracy to measure an attacker's proximity to the correct answer, i.e., the fraction of traces for which the correct website $y_i$ appears among the $k$ highest-scoring predictions:
\[
\text{Top-}k\ \text{Accuracy} = \frac{1}{N} \sum_{i=1}^{N} 
\mathbf{1}\!\left[\, y_i \in \operatorname{TopK}(p_i, k) \,\right],
\]
where $\mathbf{1}[\cdot]$ is the indicator function and $p_i$ is the scoring prediction.
A score of $0.9$ means the attacker correctly identifies the correct page within their top $k$ predicted candidates in $90\%$ of the cases.

\noindent\fbox{\textbf{\benchkey3}} \textbf{Theoretical Attacker Uncertainty under Optimal Conditions:} \emph{How much uncertainty does the defense induce in an ideal attacker’s predictions?}

\noindent Fingerprinting defenses can be tuned to evade known ML techniques, creating a false sense of security --- that is, \benchkey 1--2 remain tightly coupled to specific fingerprinting models and may fail to capture the full leakage potential present in the traffic traces.
To overcome this limitation, we measure \emph{the model-agnostic uncertainty} --- not just classification errors --- using information-theoretic methods directly on the datasets. We characterize the defense’s impact on the attackers through the prior and the conditional entropies:

\[
\begin{aligned}
    H(Y) = -\sum_{y} p(y)\,\log_{2} p(y), & &
    H(Y \mid X) = \mathbb{E}_{x}\!\left[ H\!\big(P(Y \mid X{=}x)\big) \right].
\end{aligned}
\]

\noindent where $X$ is the (defense-processed) traffic representation and $Y$ the page label. $H(Y)$ quantifies the uncertainty about the webpage before observing the traffic, while $H(Y\!\mid X)$ measures the uncertainty about webpage $Y$ after observing the features $X$ (model agnostic). 
Using these entropies, the \emph{mutual information} (MI) between the traffic representation $X$ and the label $Y$ (with $C$ unique websites) is:
\[
I(X;Y) = H(Y) - H(Y\!\mid X) \in [0,\log_2 C]~\text{bits}.
\]
Unlike accuracy, $I(X;Y)$ also credits near misses: if a model consistently narrows the candidate set without identifying the exact page, $I$ still increases. Thus, $I$ measures how much uncertainty an optimal attacker could remove, independently of any specific classifier. 
\ifextver
Two well-established yet complementary MI estimators for WF are:

\noindent\textbf{(a) WeFDE} \cite{DBLP:conf/ccs/LiGH18} estimates the mutual information using manually selected features and kernel density estimators (KDE).
WeFDE maps each trace to a vector of handcrafted features $\phi(x)$ (the same features as for k-FP --- IAT, packet, burst statistics, CUMUL), then estimates class-conditional and marginal densities with kernel density estimation (KDE) on low-dimensional feature groups. Plugging those into the MI identity yields:
\[
\widehat{I}_{\text{WeFDE}} \approx \frac{1}{N}\sum_{i=1}^{N}
\log_{2}\!\left(
\frac{\widehat{p}\big(\phi(x_i),y_i\big)}{\widehat{p}\big(\phi(x_i)\big)\,\widehat{p}(y_i)}
\right),
\]
where $\widehat{p}(\phi,y)$ and $\widehat{p}(\phi)$ are KDE estimates.  
Intuitively, the estimator approximates the relationship between the joint probability of observing a feature set $\phi$ together with a webpage label $y$ ($\widehat{p}(\phi, y)$) and the product of their marginal probabilities $\widehat{p}(\phi)$ and $\widehat{p}(y)$. 
If the features are independent of the webpage, the joint probability equals the product of the marginal probabilities, yielding $\log_{2}(1)=0$. 
Otherwise, dependence between the features and the webpage yields a nonzero value, indicating higher dependence between the features and the labels.
This produces a total MI estimate in bits and per-feature/group leakage, pinpointing what still leaks.

\noindent\textbf{(b) DeepSE-WF} \cite{DBLP:journals/popets/VeichtRB23} leverages deep learning-derived latent spaces combined with specialized $k$-NN estimators to approximate MI. 
While WeFDE provides an interpretable method for measuring information leakage directly from raw data, its accuracy depends heavily on the quality of the selected input features, and poor choices can create a false sense of security.
To better approximate information leakage, the framework trains a deep neural network to project website traces into a continuous latent representation $Z = f(X)$. It then estimates $I(Z;Y)$ using a $k$-NN-based estimator \cite{ross2014mutual}, which is specifically designed for mixtures of discrete labels ($Y$) and continuous features ($Z$):
\[
\hat{I}(Z; Y) = \psi(N) - \frac{1}{N}\sum_{i=1}^{N}\psi(n_i) + \psi(k) - \frac{1}{N}\sum_{i=1}^{N}\psi(m_i),
\]
where $\psi$ is the digamma function, $N$ is the total number of samples, $n_i$ is the number of samples sharing the class label of sample $i$, $k$ is a hyperparameter (set to $5$ in this work), and $m_i$ is the number of points \emph{regardless of class} lying within distance $\varepsilon_i$ of $x_i$, where $\varepsilon_i$ is the distance from $x_i$ to its $k$-th nearest \emph{same-class} neighbor. Finally, $\widehat{I}_{\text{DeepSE}} = \max_f \hat{I}(f(X); Y)$.
This geometric approach allows dimensionality-dependent parameters to cancel out, making the estimator more robust in high-dimensional latent spaces. 
However, by the data processing inequality \cite{cover1999elements}, $I(Z;Y) \le I(X;Y)$, so $\hat{I}(Z;Y)$ is a \emph{lower bound} on the true leakage --- meaning the actual information available to an adversary may be larger, and any privacy guarantee implied by $\hat{I}(Z;Y)$ is therefore optimistic.

\else
Two complementary MI estimators for WF are:
(a) WeFDE~\cite{DBLP:conf/ccs/LiGH18} maps each trace to handcrafted traffic features (IAT, packet and burst statistics, and CUMUL) and uses kernel density estimation (KDE) to estimate their dependence on the webpage label. 
Intuitively, if the feature distributions are independent of the webpage, the estimated leakage approaches zero; stronger page-specific dependence indicates greater leakage.
(b) DeepSE-WF~\cite{DBLP:journals/popets/VeichtRB23} instead trains a neural network to project traces into a learned latent representation and estimates its MI with the webpage label using a $k$-nearest-neighbor ($k$-NN) estimator. 
This reduces dependence on manually selected features, but by the data-processing inequality its estimate may underestimate the true leakage.
Extended estimator details are provided in the full paper version~\cite{XXX}.
\fi

\noindent\textit{Interpreting the uncertainty estimators.}
To make the privacy implications more interpretable, we express conditional entropy as an effective number of plausible webpage labels, $2^{H(Y|X)}$.
Larger values indicate greater residual uncertainty.
Since the true mutual information $I(X;Y)$ is unknown, we use the larger of the two estimated leakage values as a proxy and derive the estimator-based uncertainty measure $\mathcal{K}^{*}$:
\[
\mathcal{K}^{*} = 2^{H(Y) - \max\{\,\widehat{I}_{\text{WeFDE}},\;
\widehat{I}_{\text{DeepSE}}\}}.
\]
Here, $\max\{\widehat{I}_{\text{WeFDE}},\widehat{I}_{\text{DeepSE}}\}$ is used as a proxy for the unknown $I(X;Y)$; consequently, $\mathcal{K}^{*}$ is an estimator-derived quantity rather than a bound on the true $2^{H(Y|X)}$. 
If the estimators underestimate the true leakage, the true effective candidate-set size may be smaller than the reported $\mathcal{K}^{*}$.
The mutual-information estimates and the derived $\mathcal{K}^{*}$ values are reported as point estimates, as the employed estimators do not provide confidence intervals.

\noindent\fbox{\textbf{\benchkey4}} \textbf{Defense Operational Cost:} \emph{What is the added latency and traffic volume by the defense?}

\noindent For a webpage $y \in \mathcal{Y}$ under defense $d$, let $\text{Up}_d(y)$ denote the total upload bytes (client $\to$ server), $\text{Down}_d(y)$ the total download bytes (server $\to$ client), and $T_d(y)$ the duration until the last real content packet. 
Baseline values without defense are $\text{Up}_b(y)$, $\text{Down}_b(y)$, and $T_b(y)$. To quantify overheads, we compute the relative per-page increase ratio:
\[
    \Delta M(y) = \frac{M_d(y) - M_b(y)}{M_b(y)}, \quad M \in \{\text{Up}, \text{Down}, T\}.
\]

Similar to prior work~\cite{DBLP:conf/uss/SmithDMP22}, (1) we report the median (Q2) and the first and third quantiles (Q1--Q3) of $\Delta M$ across all pages, pooled across all case studies; 
and (2) the latency overhead excludes trailing defensive traffic after the last real content packet, as this does not affect the user's perceived page load time.

\subsubsection*{\textbf{Takeaways}}

\label{subsection:takeawayBlueprint} All experiments in the following section use identical preprocessing and evaluation steps, ensuring that score differences arise solely from the traffic itself. 
Guided by our blueprint, we can:
\begin{itemize}[left=0pt,noitemsep]
    \item Calibrate each defense per dataset to identify a practical privacy--overhead operating point (\benchkey0);
    \item Detect when a dataset is easy to fingerprint (\benchkey1) or is trivial to get close to the real answer (\benchkey2);
    \item Identify cases where current ML models underperform, compared to the actual data leakage (\benchkey3);
    \item Compare manual-feature learning against deep-learning, keeping the stronger attacker in each case (\benchkey1–2: k-FP vs. VarCNN; \benchkey3: WeFDE vs. DeepSE);
    \item Uncover design flaws (or bugs) that inflate overhead, that falsely appear beneficial (\benchkey4).
    
\end{itemize}

%% file: sections/4_method_experiments/b2_intro_defenses.tex
\section{Modelling WF Defenses in \httptwo}
\label{section:wf_defenses_emulation}

We now discuss how to model existing fingerprinting defenses using \httptwo. We first review established defensive techniques (\Cref{subsection:wf_defensive_techniques}). We then highlight \httptwo features and how they can enable defenses (\Cref{subsection:http2_background}). We finally tie these threads of knowledge by emulating \httptwo application-layer defenses for clients (\Cref{subsection:baseline_fingerprinting_defenses_clients}) and for servers (\Cref{subsection:baseline_fingerprinting_defenses_servers}).

\subsection{\textbf{Fingerprinting Defenses}}
\label{subsection:wf_defensive_techniques}

WF defenses aim to disrupt site-specific bandwidth and latency patterns, using either \emph{uniformity} (making all page loads appear identical) or \emph{unpredictability} (making each page visit look random).

\subsubsection{\textbf{Fingerprinting Defenses Categories}}
Defenses can be broadly grouped into four categories:
(1) \emph{Noise traffic techniques} add dummy packets to obscure bandwidth patterns, often effective but costly in bandwidth \cite{DBLP:journals/corr/JuarezIPDW15,DBLP:conf/uss/GongW20,DBLP:conf/infocom/AbusnainaJKNM20,DBLP:journals/corr/abs-2011-13471,DBLP:conf/uss/SmithDMP22, DBLP:journals/popets/HollandH22};
(2) \emph{Padding techniques} (either through traffic morphing or adversarial perturbations)  modify packet sizes to mimic other traffic patterns, usually with lower bandwidth overhead than pure noise injection, but often leaving timing patterns intact \cite{DBLP:conf/uss/WangG17,DBLP:journals/tifs/RahmanIMW21,al-naami_bimorphing_2021,DBLP:conf/uss/NasrBH21,DBLP:journals/corr/abs-2102-04291}. 
(3) \emph{Traffic splitting techniques}, which alter the number of requests and responses \cite{DBLP:conf/ndss/LuoZCLCP11,DBLP:conf/uss/SmithDMP22,DBLP:journals/corr/JuarezIPDW15,DBLP:conf/uss/GongW20};
and (4) \emph{Traffic-shaping techniques} enforce fixed-size packets at fixed intervals, achieving strong uniformity guarantees but at the cost of higher latency and bandwidth \cite{DBLP:conf/sp/DyerCRS12,DBLP:conf/wpes/CaiNJ14,DBLP:conf/ccs/CaiNWJG14,lu_dynaflow_2018,DBLP:conf/uss/SabziVGSLM24, DBLP:journals/popets/HollandH22}.

\subsubsection{\textbf{Fingerprinting Defenses: \tor vs.\ Plain \https}}
Most fingerprinting defense research focuses on \tor traffic \cite{DBLP:journals/corr/JuarezIPDW15,DBLP:conf/uss/GongW20,DBLP:conf/sp/DyerCRS12,DBLP:conf/wpes/CaiNJ14,DBLP:conf/ccs/CaiNWJG14,lu_dynaflow_2018,DBLP:conf/uss/SabziVGSLM24, wang2026cease}, where all data is multiplexed into fixed-size cells within a single tunnel. 
Plain \https defenses instead complement \tor/\vpn solutions: they do not hide the destination domain, but aim to reduce subpage leakage.
This setting introduces additional challenges compared to the \tor/\vpn defenses: packet sizes vary, and a page load spans multiple parallel \tcp connections that can be fingerprinted independently.
Some defenses are designed specifically for \https \cite{DBLP:journals/popets/CherubinHJ17}, while others originally built for \tor can, in principle, be adapted \cite{DBLP:conf/wpes/CaiNJ14,DBLP:conf/uss/GongW20}. However, (1) traffic-shaping defenses must also address packet-size leakage and (2) unlike \tor, noise added to one connection does not propagate to the encapsulating layer --- every leaking connection must be defended individually. 

Defenses can be deployed at different points in the communication path: (1) Client-side (in the browser, OS, or client application), where traffic can be perturbed before it leaves the user’s device \cite{DBLP:conf/uss/GongW20,DBLP:journals/corr/JuarezIPDW15,DBLP:conf/uss/SmithDMP22,al-naami_bimorphing_2021,DBLP:conf/wpes/CaiNJ14,DBLP:journals/popets/CherubinHJ17,DBLP:conf/ndss/LuoZCLCP11}; or (2) Server-side (at the web server or CDN), where content delivery patterns, resource sizes, or response delays can be modified \cite{DBLP:journals/popets/CherubinHJ17,DBLP:conf/infocom/AbusnainaJKNM20}. 
In the following sections, we focus on one-sided (client-only or server-only) \httptwo application-layer defenses powered by unpredictability.

\subsection{\textbf{\httptwo Privacy Potential}}
\label{subsection:http2_background}

\httptwo represents roughly $60\%$ of browser traffic~\cite{cloudflareExaminingHTTP3} and introduces several features beyond \httpone that can be relevant to fingerprinting defenses.
\Cref{tab:http_version_comparison} outlines the key similarities and differences across \http versions. 

As defined in RFC~9113 \cite{DBLP:journals/rfc/rfc9113}, the basic unit of the \httptwo communication is a \emph{frame}. A \emph{stream} is a bidirectional sequence of frames between the client and the server, mapped to a single client request. An \httptwo \emph{connection} corresponds to a single \tcp connection and comprises multiple streams, i.e., each carrying one or more requests and responses exchanged with the same server/subdomain.

\subsubsection{\textbf{\httpone Inheritance}} 
\httptwo inherits \httpone's header semantics --- the same fields for method, path, status, and content negotiation. 
It also inherits Range requests (RFC~7233~\cite{ietf7233Hypertext}), which allow clients to fetch specific byte ranges of a resource, enabling application-layer traffic splitting. 

\subsubsection{\textbf{\httptwo Multiplexing}} 
On the transport side, \httptwo replaces \httpone pipelining with stream multiplexing: multiple requests and responses are interleaved concurrently over a single persistent \tcp connection without ordering constraints, eliminating head-of-line blocking. Beyond performance 
gains~\cite{akamaiMultiplexing}, multiplexing blurs the boundaries between individual encrypted requests and responses, with direct implications for connection privacy.

\begin{table}[t!]
\caption{Comparison between the \http versions.}
\centering
\resizebox{\columnwidth}{!}{
    \begin{tabular}{@{}cccc@{}}
    \toprule
    \textbf{Feature}      & \textbf{\httpone} & \textbf{\httptwo}           & \textbf{\httpthree} \\ \midrule
    Request Resource      & \cmark            & \cmark                      & \cmark              \\
    Range Requests        & \cmark            & \cmark                      & \cmark              \\
    Transport             & \tcp              & \tcp                        & \udp                \\
    Flow Control          & \xmark            & \cmark                      & \cmark              \\
    Concurrent Streams    & \xmark            & \cmark                      & \cmark              \\
    Stream Prioritization & \xmark            & \cmark                      & \cmark              \\
    Header Compression    & \xmark            & \cmark                      & \cmark              \\
    Server-Push           & \xmark            & (\cmark)                      & \xmark              \\
    103 Early Hints       & \cmark            & \cmark                      & \cmark              \\
    Built-in Padding      & \xmark            & Limited $\leq$ 255 B        & Unlimited           \\ 
    Connection Probing    & \xmark            & \texttt{PING} (8 B)         & \texttt{PING} (no payload)           \\
    \bottomrule
    \end{tabular}
}
\label{tab:http_version_comparison}
\end{table}

\subsubsection{\textbf{\httptwo Stream Prioritization}} In \httptwo, clients and servers can change the priority of streams depending on the content type (e.g., HTML, CSS, JS), the content location in the pages (above or below the visible fold), or from developer hints \cite{perfplanetHTTP3Prioritization}. Stream reprioritization can also lead to unpredictable burst patterns, thus improving communication privacy.

\subsubsection{\textbf{\httptwo Flow Control}}
\httptwo introduces its own flow control at the application layer, independent of \tcp's transport-layer flow control. 
While \tcp flow control governs how much data can be in-flight on the network at once, \httptwo flow control operates per-stream: the receiver advertises an initial window size for each stream and can increment it via \texttt{WINDOW\_UPDATE} frames, allowing it to pace or 
pause individual streams independently without affecting others on the same connection. 
This per-stream granularity has direct fingerprinting implications: by varying window sizes or deliberately delaying \texttt{WINDOW\_UPDATE} frames, the receiver can alter the sizes and timing of downloaded DATA frames at the \http layer, independently 
of what \tcp would otherwise permit.

\subsubsection{\textbf{\httptwo Resource Suggestions}} \httptwo servers can proactively suggest or even deliver resources to the clients.
One option is Server-Push frames, which let a server proactively send responses linked to a client’s earlier request. For instance, after an HTML request, the server can push JavaScript, CSS, or images without waiting for additional requests. However, support for Server-Push has declined in recent years, with Chrome deprecating its support in 2022. Our evaluation of Server Push-based defenses serves primarily to establish theoretical bounds on server-side defense efficacy.
A more widely adopted alternative is the \http `103 Early Hints' status code \cite{DBLP:journals/rfc/rfc8297}, which lets servers indicate critical resources for the client to fetch proactively. Unlike Server-Push, this approach gives the client full control over which resources to request, at the cost of an extra round trip.

\subsubsection{\textbf{\httptwo Built-in Padding}} \httptwo offers limited support for padding for \texttt{DATA}, \texttt{HEADERS}, and \texttt{PUSH\_PROMISE} frames, with up-to 255 bytes of extra noise. In contrast, \quic provides dedicated PADDING frames~\cite{ietf9000QUIC}, and \httpthree DATA frames can include arbitrary-length padding, making padding strategies more flexible than \httptwo’s 255-byte limit per frame.

\subsubsection{\textbf{\httptwo Connection Probing.}} \httptwo uses \texttt{PING} frames for liveness checks and RTT measurements. Each \texttt{PING} frame must contain an 8-byte payload, which the remote party will echo. 

\subsubsection{\textbf{\httptwo Opportunities for Fingerprinting Defenses}}

Unlike \quic, which offers fine-grained stream-level flow control and lets applications dictate when and how packets are sent, \httptwo relies on \tcp, where the OS stack manages transmission. As a result, defenses that depend on strict fixed-rate packet flows can only be approximated in \httptwo, since it lacks {\quic}’s frame-level control.

While \httptwo offers features that could enhance privacy, their defensive potential has not been systematically quantified. Prior studies generally found that website fingerprinting remains effective against \httptwo traffic~\cite{DBLP:journals/corr/Morla17,DBLP:conf/networking/GhietteD20,wang_high_2020,lin2019measuring,liang_tail_2022,mitra_depending_2020,DBLP:conf/IEEEares/MartinoQL19,mitra_depending_2020}, but did not actively leverage \httptwo-specific mechanisms to strengthen communication privacy. To that end, in addition to emulating established defenses with \httptwo, we are also interested in exploring the privacy impact of the aforementioned \httptwo features.

%% file: sections/4_method_experiments/c_ref_client_defenses.tex
\subsection{\httptwo Client-Side Defenses}
\label{subsection:baseline_fingerprinting_defenses_clients}

We now show how \httptwo clients can increase privacy in the \numdatasets case studies from \Cref{subsection:realworld_dataset_baselines}.
To that end, we emulate and evaluate four defensive strategies using \httptwo features: (1) HTTPOS \cite{DBLP:conf/ndss/LuoZCLCP11}, a traffic splitting technique; (2) LLaMA \cite{DBLP:journals/popets/CherubinHJ17}, a traffic morphing defense; (3) FRONT \cite{DBLP:conf/uss/GongW20}, a noise insertion strategy; and (4) Tamaraw  \cite{DBLP:conf/ccs/CaiNWJG14}, a traffic-shaping defense emulated from the \emph{client-side}. The following describes how each defense is adapted to \httptwo and includes a state machine diagram showing its page-load behavior.


\subsubsection{\textbf{HTTPOS}} HTTPOS \cite{DBLP:conf/ndss/LuoZCLCP11} obscures traffic patterns by splitting requests using the Range header to fetch specific or overlapping parts of binary resources (e.g., images), and by constraining the receive window so that responses arrive in smaller units. The original defense manipulates the \tcp advertised window; we approximate this behavior using \httptwo flow control, which constrains DATA delivery while leaving the frame-to-segment mapping to the kernel. Request splitting changes the number of transmitted units, while overlapping ranges change how many bytes are transferred.

\begin{figure}[t!]
        \centering
        \includegraphics[width=\columnwidth]{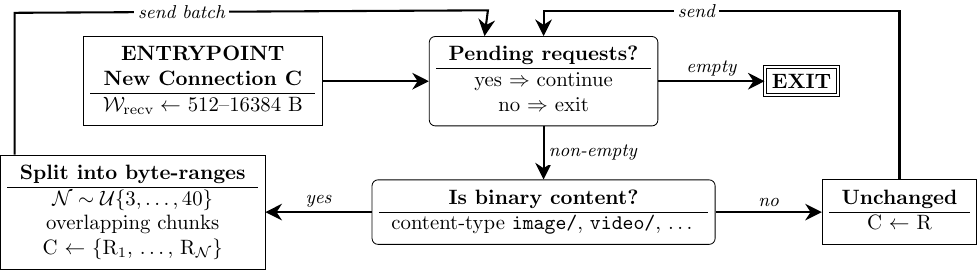}	
	    \caption{\httptwo emulation of the HTTPOS Defense \cite{DBLP:conf/ndss/LuoZCLCP11}. }
        \Description{\httptwo emulation of the HTTPOS Defense \cite{DBLP:conf/ndss/LuoZCLCP11}. }
        \label{fig:client_defense_emulation_httpos}
\end{figure}

Yet, with this approach, the Range header is mainly applicable to binary data, and servers may ignore it altogether --- e.g., by returning the full image even if only a byte range is requested. In our experiments, we assume the server cooperates.
\Cref{fig:client_defense_emulation_httpos} shows the \httptwo version of the defense.

\ifextver
\begin{figure}[b]
    \centering
    \includegraphics[width=\columnwidth]
    {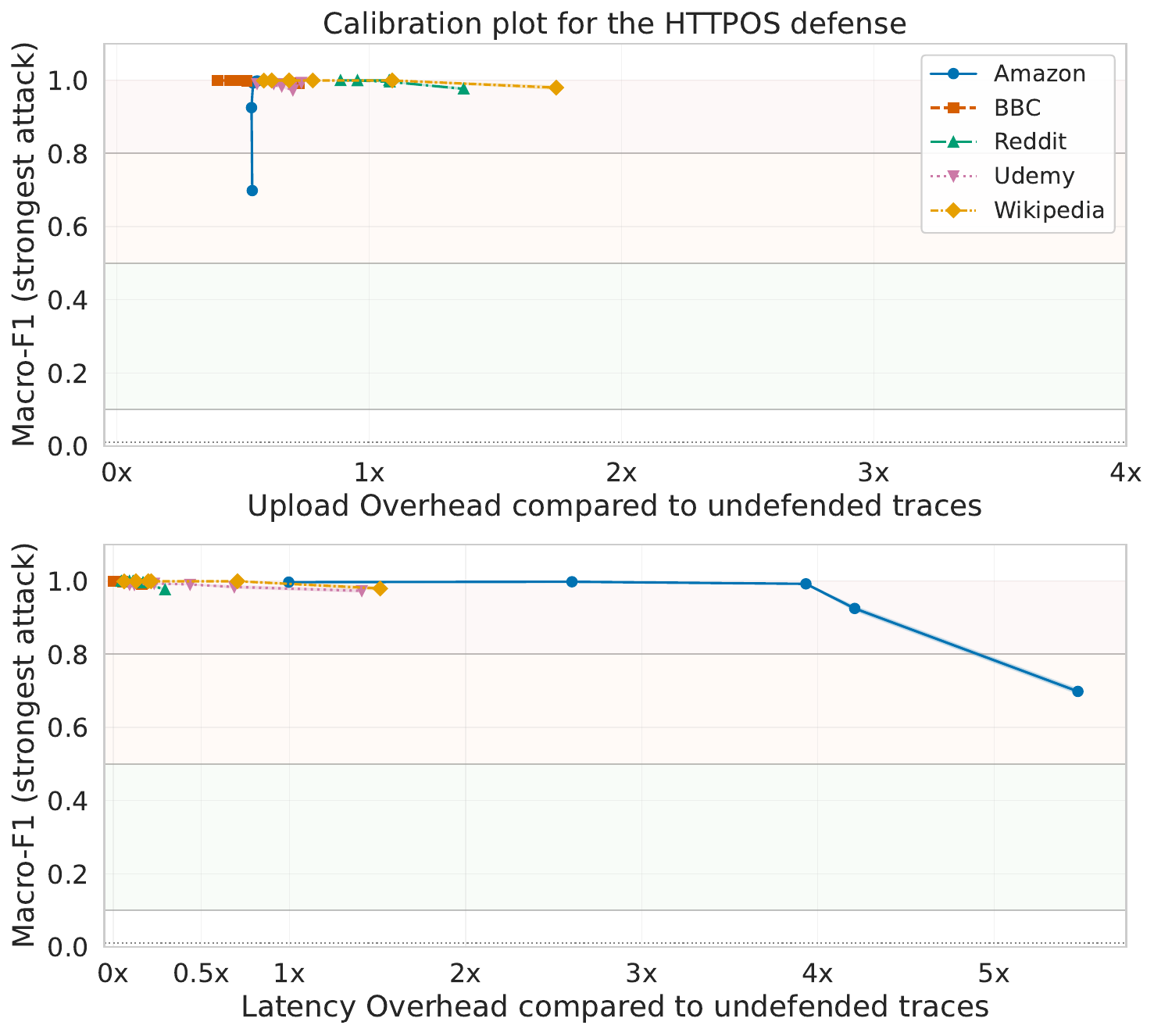}
    \caption{Calibration for the HTTPOS defense (\benchkey0).
    Strongest-attacker Macro-F1 per dataset against bandwidth (top) and latency (bottom) overhead. Horizontal lines mark the weak, moderate, and strong attacker thresholds; the dotted line is random guessing.}
    \Description{Calibration results for HTTPOS showing strongest-attacker Macro-F1 against bandwidth and latency overhead across the evaluated datasets and defense configurations.}
    \label{fig:client_defense_calibration_httpos}
\end{figure}
\fi

For the HTTPOS defense calibration (\textbf{\benchkey0}), we jointly vary the initial \httptwo flow-control window $\mathcal{W}$ and the lower and upper limits $[S_{\min},S_{\max}]$ of the range-split count $\mathcal{N}$, with $\mathcal{N}\sim\mathcal{U}\{S_{\min},\ldots,S_{\max}\}$, across six candidate configurations spanning $\mathcal{W}=16384$--$512$\,B and range-split bounds from $[3,5]$ to $[20,40]$.
Compared to the original implementation, we additionally vary the initial flow-control window.
\ifextver
The six candidate configurations use $\mathcal{W}\in\{16384,\allowbreak 8192,\allowbreak 4096,\allowbreak 2048,\allowbreak 1024,\allowbreak 512\}$\,B with corresponding range-split bounds $[S_{\min},S_{\max}]\allowbreak \in\allowbreak \{[3,5],\allowbreak [4,7],\allowbreak [5,8],\allowbreak [5,10],\allowbreak [10,20],\allowbreak [20,40]\}$.
\Cref{fig:client_defense_calibration_httpos} shows that HTTPOS provides limited protection across the calibration sweep.
On Amazon, increasing the defense intensity reduces the calibration Macro-F1 to $0.699$ for $\mathcal{W}=1024$\,B and $\mathcal{N}\sim\mathcal{U}\{10,\ldots,20\}$.
On BBC and Reddit, the calibration Macro-F1 remains above $0.97$ under the same configuration, and on Udemy and Wikipedia it remains above $0.97$ even with $\mathcal{W}=512$\,B and $\mathcal{N}\sim\mathcal{U}\{20,\ldots,40\}$.
These results indicate that more aggressive range splitting alone does not substantially suppress fingerprinting leakage on most datasets.
\fi
Following calibration, we therefore retain $\mathcal{W}=1024$\,B with $\mathcal{N}\sim\mathcal{U}\{10,\ldots,20\}$ for Amazon, BBC, and Reddit, and $\mathcal{W}=512$\,B with $\mathcal{N}\sim\mathcal{U}\{20,\ldots,40\}$ for Udemy and Wikipedia.

\begin{figure}[b!]
        \centering
        \includegraphics[width=\columnwidth]{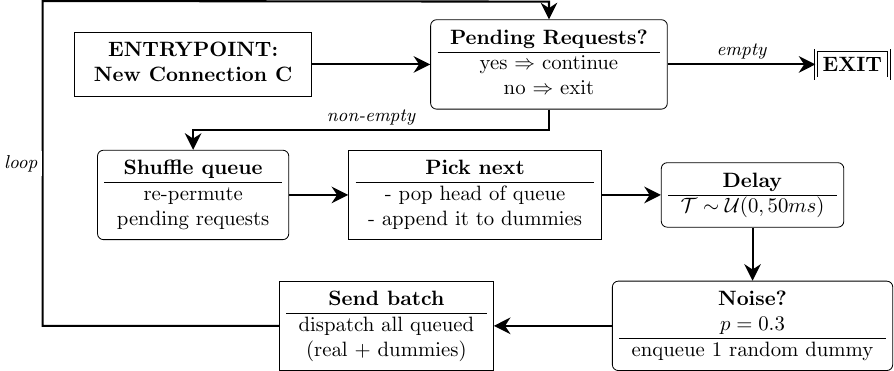}	
	    \caption{\httptwo emulation of the LLaMA Defense \cite{DBLP:journals/popets/CherubinHJ17}. }
    	\Description{\httptwo emulation of the LLaMA Defense \cite{DBLP:journals/popets/CherubinHJ17}. }
        \label{fig:client_defense_emulation_llama}
\end{figure}

\subsubsection{\textbf{LLaMA}} LLaMA \cite{DBLP:journals/popets/CherubinHJ17} is an application-layer defense that obfuscates traffic by randomizing request order, using \http pipelining, introducing random delays, and injecting dummy traffic.
In \httptwo, request pipelining is natively supported through stream multiplexing. Each stream can be delayed independently without stalling the entire connection.
LLaMA can leverage these features by grouping requests into random batches (up to 5 in our implementation) and reordering them before sending.
Every request is delayed by $\mathcal{T} \sim \mathcal{U}(0, \tau)$; the original draws $\tau$ as half the median page load time, which assumes concurrent dispatch, so we scale it by the mean number of requests per page to obtain the same latency budget under sequential replay.
To further increase ambiguity, LLaMA probabilistically issues dummy \http requests after each real request and received response, injecting noise into the traffic pattern.
\Cref{fig:client_defense_emulation_llama} illustrates the \httptwo version of the defense.

For the LLaMA defense calibration (\textbf{\benchkey0}), we vary the dummy-request probability over $p_d\in\{0,0.15,0.25,0.30,0.50,0.80\}$ while keeping the dataset-specific delay bound $\tau$ fixed.
\ifextver
\begin{figure}[t]
    \centering
    \includegraphics[width=\columnwidth]
    {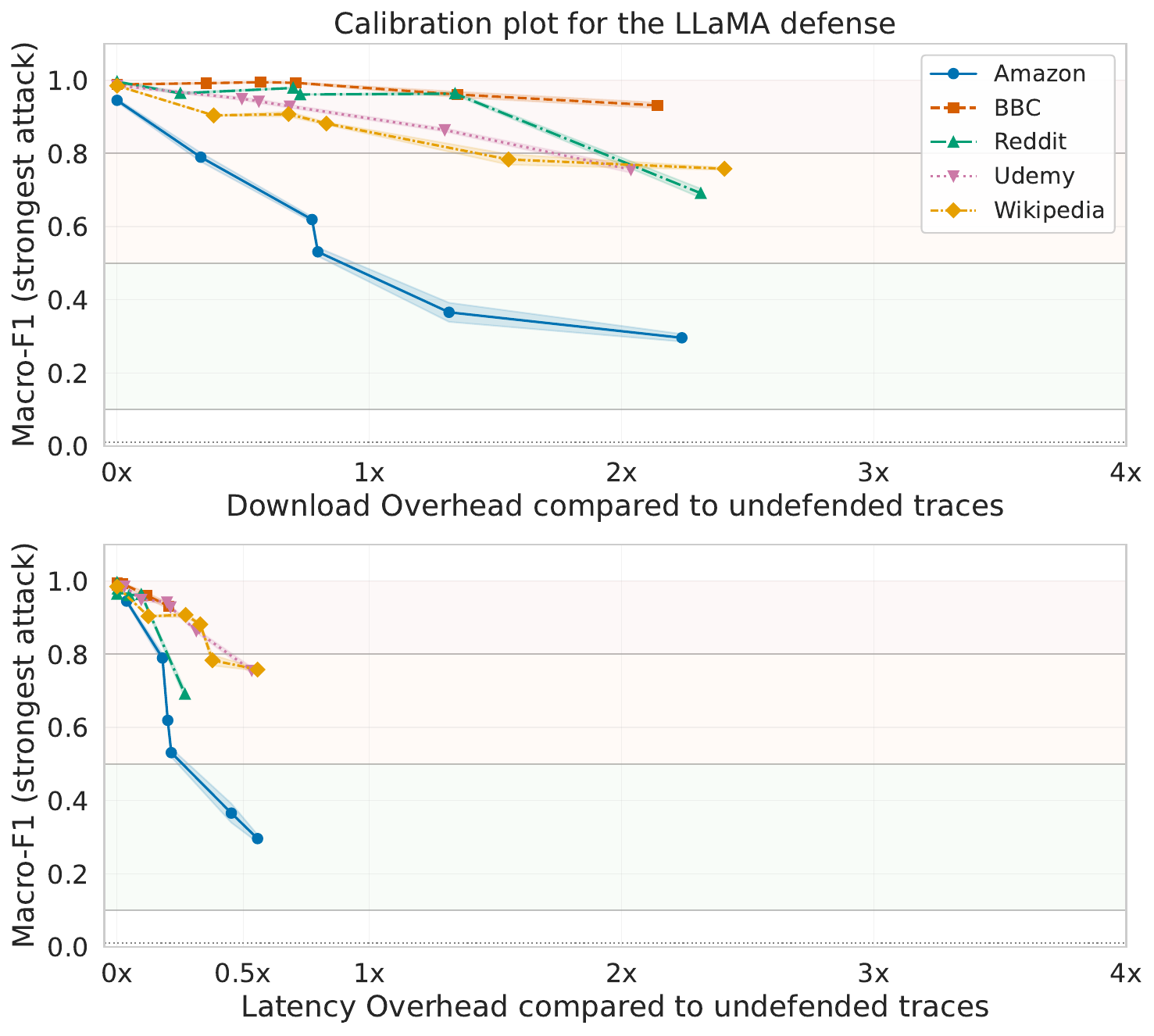}
    \caption{Calibration for the LLaMA defense (\benchkey0). Strongest-attacker Macro-F1 per dataset against bandwidth (top) and latency (bottom) overhead. Horizontal lines mark the weak, moderate, and strong attacker thresholds; the dotted line is random guessing.}
    \Description{Calibration results for LLaMA showing strongest-attacker Macro-F1 against bandwidth and latency overhead across the evaluated datasets and dummy-request probabilities.}
    \label{fig:client_defense_calibration_llama}
\end{figure}
\Cref{fig:client_defense_calibration_llama} shows that LLaMA generally benefits from increasing the dummy-request probability, but the magnitude of the improvement is strongly dataset-dependent.
On Amazon, the calibration Macro-F1 decreases to $0.296$ at $p_d=0.8$, while on BBC, Reddit, and Udemy the same configuration reaches $0.931$, $0.691$, and $0.756$, respectively.
On Wikipedia, $p_d=0.5$ already reaches $0.784$, while increasing $p_d$ to $0.8$ improves calibration Macro-F1 only to $0.758$ while increasing both bandwidth and latency overhead.

\fi
The calibration shows that LLaMA generally benefits from increasing the dummy-request probability, but the magnitude of the improvement is strongly dataset-dependent.
Consequently, we select $p_d=0.8$ for Amazon, BBC, Reddit, and Udemy, and $p_d=0.5$ for Wikipedia.

\begin{figure}[b!]
    \centering
    \includegraphics[width=\columnwidth]{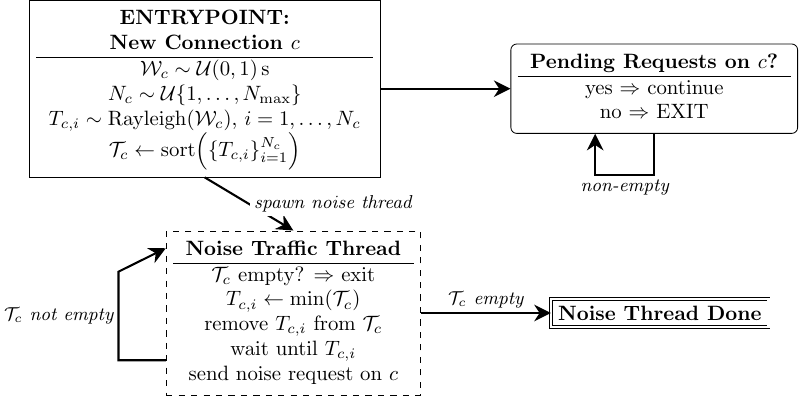}
    \caption{\httptwo emulation of the FRONT defense~\cite{DBLP:conf/uss/GongW20}.}
    \Description{\httptwo emulation of the FRONT defense~\cite{DBLP:conf/uss/GongW20}.}
    \label{fig:client_defense_emulation_front}
\end{figure}

\subsubsection{\textbf{FRONT}}
The FRONT defense~\cite{DBLP:conf/uss/GongW20} injects $N$ dummy packets at timestamps drawn from a $\mathrm{Rayleigh}(\mathcal{W})$ distribution, where the scale $\mathcal{W}\sim\mathcal{U}(0,1)$ is randomized per connection. 
In our \httptwo adaptation, for each connection $c$, we sample the number of scheduled dummy transmissions as $N_c\sim\mathcal{U}\{1,\ldots,N_{\max}\}$, independently sample $\mathcal{W}_c\sim\mathcal{U}(0,1)$ seconds, and draw $N_c$ transmission times from $\mathrm{Rayleigh}(\mathcal{W}_c)$.
Thus, both the dummy volume and timing pattern vary across connections.
QCSD~\cite{DBLP:conf/uss/SmithDMP22} adapted FRONT to \quic, and the same noise-scheduling principles apply to \httptwo.
\Cref{fig:client_defense_emulation_front} summarizes our \httptwo emulation and its parameters.

For the FRONT defense calibration (\textbf{\benchkey0}), we vary the maximum per-connection dummy-stream budget $N_{\max} \in \{20,\allowbreak 50,\allowbreak 110,\allowbreak 200,\allowbreak 350,\allowbreak 500\}$.
\ifextver
\begin{figure}[t]
    \centering
    \includegraphics[width=\columnwidth]
    {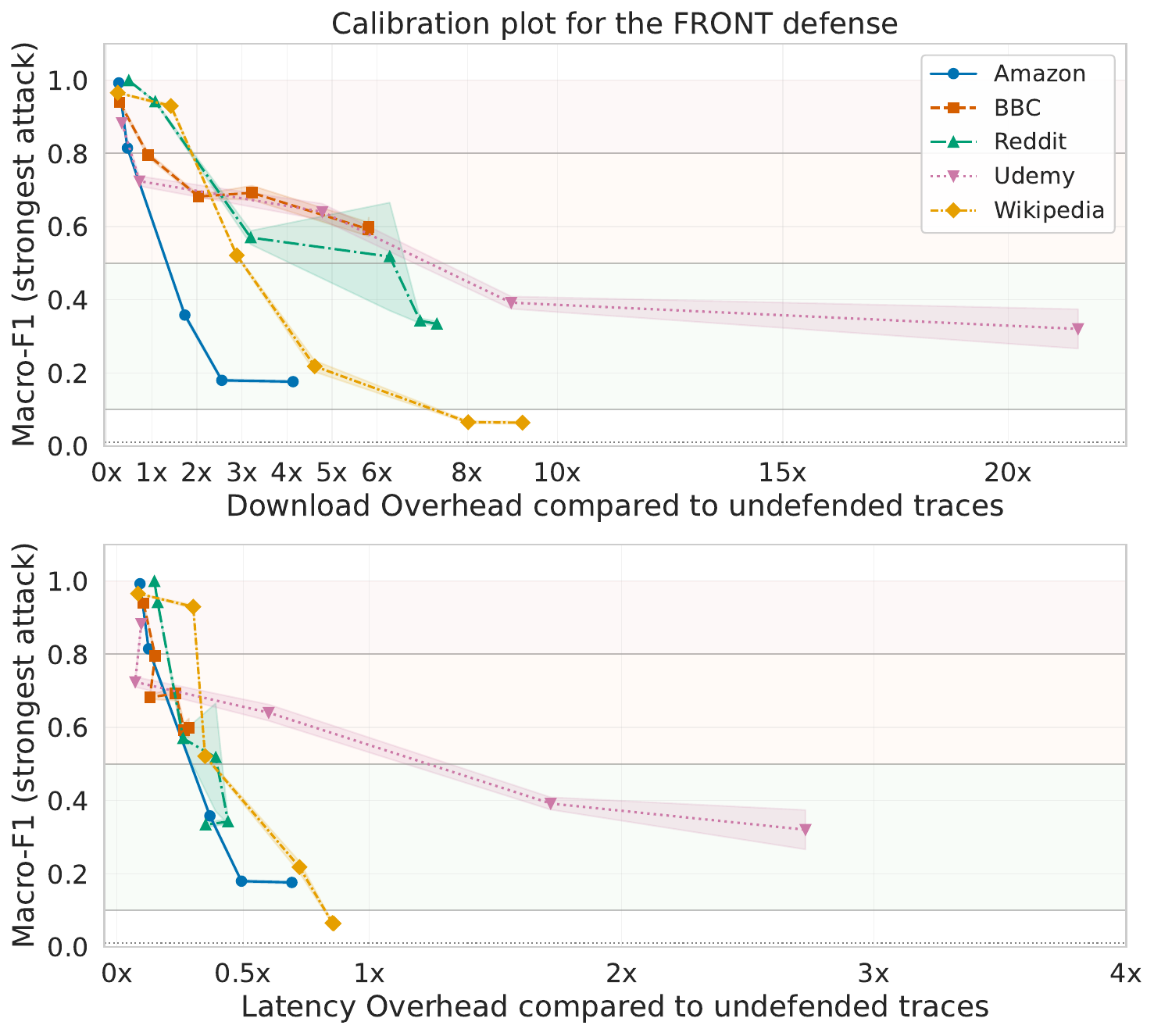}
    \caption{Calibration for the FRONT defense (\benchkey0).
    Strongest-attacker Macro-F1 per dataset against bandwidth (top) and latency (bottom) overhead. Horizontal lines mark the weak, moderate, and strong attacker thresholds; the dotted line is random guessing.}
    \Description{Calibration results for FRONT showing strongest-attacker Macro-F1 against bandwidth and latency overhead across the evaluated datasets and dummy-stream budgets.}
    \label{fig:client_defense_calibration_front}
\end{figure}
\Cref{fig:client_defense_calibration_front} shows a clear privacy--overhead knee on all five datasets.
On Amazon, the calibration Macro-F1 decreases from $0.358$ at $N_{\max}=110$ to $0.180$ at $N_{\max}=200$, whereas increasing the budget to $350$ provides almost no further improvement ($0.176$) while substantially increasing bandwidth overhead.
Similarly, Reddit improves from $0.518$ at $N_{\max}=200$ to $0.343$ at $N_{\max}=350$, while increasing the budget to $500$ reaches only $0.334$; Wikipedia reaches $0.065$ at $N_{\max}=350$ and $0.064$ at $N_{\max}=500$.
On BBC, $N_{\max}=350$ achieves the lowest observed Macro-F1 ($0.593$), with no benefit from increasing the budget to $500$.
On Udemy, increasing $N_{\max}$ from $200$ to $350$ improves the calibration Macro-F1 from $0.392$ to $0.320$, but increases downstream and upstream overhead from $8.97$ and $7.16$ to $21.55$ and $16.21$, respectively; we therefore retain the lower-overhead configuration.

\fi
The calibration shows that FRONT exhibits a clear privacy--overhead knee on all five datasets, with the selected operating points yielding a calibration Macro-F1 of at most $0.593$.
We therefore select $N_{\max}=200$ for Amazon and Udemy, and $N_{\max}=350$ for BBC, Reddit, and Wikipedia.

\subsubsection{\textbf{CL-Tamaraw}} 
\begin{figure}[t]
        \centering
        \includegraphics[width=\columnwidth]{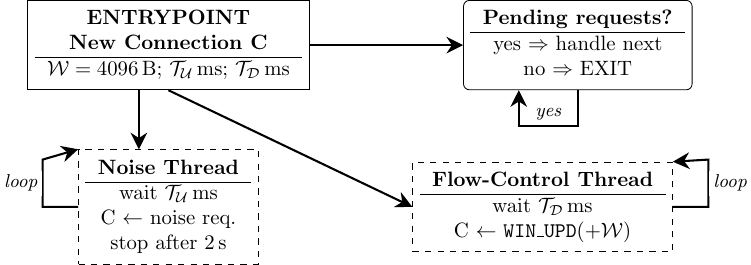}	
        \caption{\httptwo emulation of the CL-Tamaraw Defense \cite{DBLP:conf/ccs/CaiNWJG14}.  }
        \Description{\httptwo emulation of the CL-Tamaraw Defense \cite{DBLP:conf/ccs/CaiNWJG14}.  }
        \label{fig:client_defense_emulation_tamaraw}        
\end{figure}

The Tamaraw defense \cite{DBLP:conf/ccs/CaiNWJG14} obfuscates traffic by enforcing fixed-size packets at a constant rate. While feasible in \tor\ entry nodes --- where traffic can be modified bidirectionally --- this makes Tamaraw difficult to adapt to \httptwo.
QCSD \cite{DBLP:conf/uss/SmithDMP22} showed that similar behavior can be reproduced in \quic using flow control and multiplexing.
\httptwo frame sizes are under application control, but the mapping of frames to \tcp segments is governed by the kernel, so a fixed-rate fixed-size packet stream cannot be enforced from userspace; instead, we approximate constant-rate behavior through flow-control window updates --- i.e., the client can request a small initial flow-control window, thereby limiting the size of DATA frames the server can send. After receiving a fixed amount of data, the client can intentionally delay sending window update frames --- pausing the server’s transmission --- to mimic fixed-rate response defenses.  
Due to the \tcp flow control in the kernel, this does not achieve a perfectly constant rate; yet, it still alters the observable timing patterns of the response. In addition, the defense issues noise requests at a fixed rate, further disrupting burst and cumulative statistics.
\Cref{fig:client_defense_emulation_tamaraw} outlines our client-side emulation of Tamaraw using \httptwo, noting that it is not a full replication of the original \tor\ defense.

\ifextver
\begin{figure}[b]
    \centering
    \includegraphics[width=\columnwidth]
    {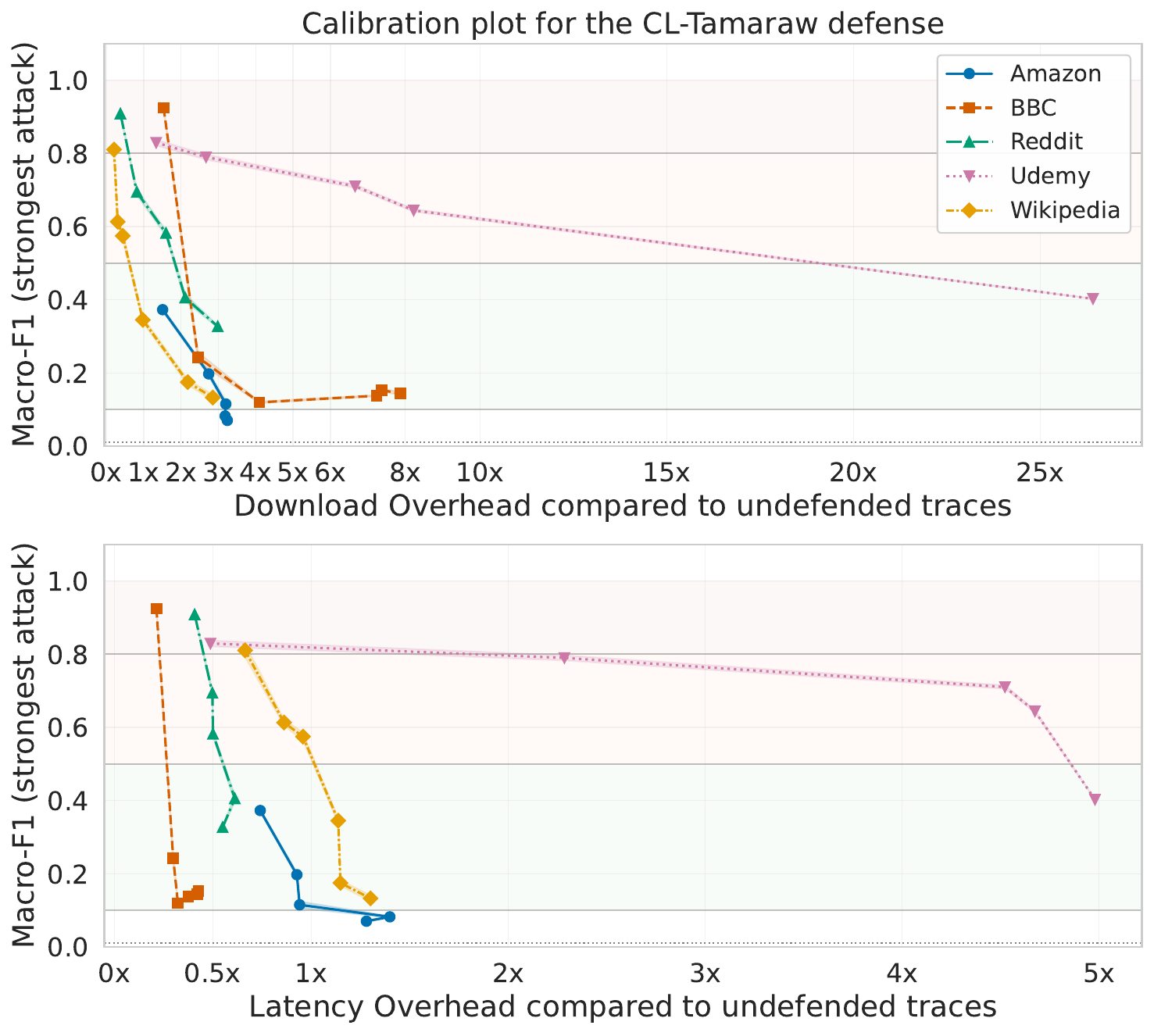}
    \caption{Calibration for the CL-Tamaraw defense (\benchkey0).
    Strongest-attacker Macro-F1 per dataset against bandwidth (top) and latency (bottom) overhead. Horizontal lines mark the weak, moderate, and strong attacker thresholds; the dotted line is random guessing.}
    \Description{Calibration for the CL-Tamaraw defense (\benchkey0).}
    \label{fig:client_defense_calibration_cltamaraw}
\end{figure}
\fi
For the CL-Tamaraw defense calibration (\textbf{\benchkey0}), we vary only the dummy-request interval $\mathcal{T}_U \in \{320,\allowbreak 80,\allowbreak 40,\allowbreak 20,\allowbreak 10,\allowbreak 5\}$\,ms, while holding the flow-control window at $\mathcal{W}=4096$\,B and the receive-delay bound $\mathcal{T}_D$ at $10$\,ms. The receive-delay threshold is fixed to $4096$\,B.
\ifextver
\Cref{fig:client_defense_calibration_cltamaraw} shows that CL-Tamaraw exhibits strong dataset dependence and early saturation on several datasets.
On Amazon, the calibration Macro-F1 reaches $0.082$ at $\mathcal{T}_U=20$\,ms and improves only to $0.070$ at $10$\,ms, so we select $20$\,ms.
On BBC, $\mathcal{T}_U=40$\,ms already achieves the lowest observed Macro-F1 ($0.120$); shorter intervals increase overhead without improving protection, so $40$\,ms is selected.
Reddit continues to benefit from stronger shaping throughout the evaluated range, reaching $0.327$ at $\mathcal{T}_U=5$\,ms, which is therefore selected.
Udemy presents a substantially less favorable trade-off: reducing $\mathcal{T}_U$ from $20$\,ms to $5$\,ms reduces the calibration Macro-F1 from $0.644$ to $0.403$, but increases downstream overhead from $8.23$ to $26.41$ and upstream overhead from $10.47$ to $28.29$.
We therefore retain $20$\,ms as the practical operating point.
Finally, Wikipedia reaches $0.175$ at $\mathcal{T}_U=10$\,ms, while $5$\,ms provides only a modest additional reduction to $0.133$; we select $10$\,ms.

\fi
Using a calibration target of strongest-attacker Macro-F1 closest below $0.5$, while avoiding unnecessarily costly configurations, we select $\mathcal{T}_U=20$\,ms for Amazon and Udemy, $40$\,ms for BBC, $5$\,ms for Reddit, and $10$\,ms for Wikipedia.

\subsubsection{\textbf{Fingerprinting on Client-Defended Datasets}} 
\Cref{tab:client_defenses_realworld_f1_score} summarizes the Macro-F1, Top-5 and $\mathcal{K}^{*}$ scores for each dataset under calibrated client-side defenses, using the strongest attacker for each dataset--defense pair, with attackers hyperparameter-tuned independently for each pair (\textbf{\benchkey1 -- 3}).
CL-Tamaraw provides the strongest resilience across all five datasets.

We first look at the Macro-F1 score. 
FRONT provides the second strongest protection overall, with its best results on Amazon ($0.54$) and Wikipedia ($0.43$), but remains much less effective on BBC, Reddit, and Udemy ($0.89$, $0.88$, and $0.89$, respectively).
CL-Tamaraw achieves the lowest Macro-F1 on every dataset, reaching $0.24$ on Amazon and BBC, $0.38$ on Reddit, and $0.42$ on Wikipedia, but remains weak
on Udemy ($0.88$).
In contrast, HTTPOS is largely ineffective across all five datasets, with Macro-F1 ranging from $0.86$ to $0.99$.
LLaMA similarly fails to provide meaningful protection, with Macro-F1 remaining above $0.80$ for all datasets.
From the attacker's perspective, RobustFP-CNN and Holmes shine against heavier defenses (FRONT, CL-Tamaraw, LLaMA), while k-FP dominates against lighter defenses (HTTPOS). 
One explanation is that defended traces are longer (noise packets extend the sequence), giving the CNN models more context windows to find stable patterns that survive noise injection.

The Top-5 results further distinguish defenses that merely disrupt the attacker's top-ranked prediction from those that induce broader ranking uncertainty. 
With CL-Tamaraw, Top-5 accuracy falls to under $0.55$ on Amazon, BBC, and Reddit, respectively, indicating that the true webpage is absent from even the attacker's five highest-ranked candidates in roughly half of the traces. 
Wikipedia shows a similar, though weaker, effect ($0.62$). 
In contrast, Udemy remains highly identifiable under every defense: even with CL-Tamaraw, Top-5 accuracy is $0.99$. 
FRONT exhibits meaningful Top-5 robustness primarily on Wikipedia ($0.70$), while its Top-5 accuracy remains at least $0.87$ on the other datasets.

\begin{table}[t]
\centering
\caption{Baseline resilience of the \httptwo \emph{Client Defenses} against the best-performing evaluated attacker (\benchkey1 and 2), together with the estimator-derived anonymity-set proxy $\mathcal{K}^{*}$ (\benchkey3). Macro-F1 and Top-5 are means; their 95\% CIs are below 0.02 and omitted.}
\label{tab:client_defenses_realworld_f1_score}
    \setlength{\tabcolsep}{3pt}

    \begin{tabular}{@{}c@{\,}c@{\,}cccc@{}}
    \toprule
    \multicolumn{1}{c}{\textbf{Dataset}} & \textbf{Metric} & \multicolumn{1}{c}{\textbf{HTTPOS}} & \multicolumn{1}{c}{\textbf{LLaMA}} & \multicolumn{1}{c}{\textbf{FRONT}} & \multicolumn{1}{c}{\textbf{CL-TAM}} \\ \midrule
    \multirow{3}{*}{Amazon} & Macro-F1          & $0.86$ & $0.82$ & $0.54$  & $\mathbf{0.24}$ \\
                            & Top-5             & $0.96$ & $0.96$ & $0.87$  & $\mathbf{0.51}$ \\
                            & $\mathcal{K}^{*}$ & $2.84$ & $3.62$ & $16.80$ & $\mathbf{38.43}$ \\ \midrule
    \multirow{3}{*}{BBC}    & Macro-F1          & $0.99$ & $0.99$ & $0.89$  & $\mathbf{0.24}$ \\
                            & Top-5             & $1.00$ & $1.00$ & $0.95$  & $\mathbf{0.54}$ \\
                            & $\mathcal{K}^{*}$ & $1.10$ & $1.13$ & $2.28$  & $\mathbf{26.64}$ \\ \midrule
    \multirow{3}{*}{Reddit} & Macro-F1          & $0.98$ & $0.97$ & $0.88$  & $\mathbf{0.38}$ \\
                            & Top-5             & $1.00$ & $0.99$ & $0.98$  & $\mathbf{0.47}$ \\
                            & $\mathcal{K}^{*}$ & $1.22$ & $1.31$ & $2.57$  & $\mathbf{34.08}$ \\ \midrule
    \multirow{3}{*}{Udemy}  & Macro-F1          & $0.99$ & $0.96$ & $0.89$  & $\mathbf{0.88}$ \\
                            & Top-5             & $1.00$ & $1.00$ & $1.00$  & $\mathbf{0.99}$ \\
                            & $\mathcal{K}^{*}$ & $1.10$ & $1.22$ & $1.82$  & $\mathbf{2.48}$ \\ \midrule
    \multirow{3}{*}{Wiki}   & Macro-F1          & $0.99$ & $0.98$ & $0.43$  & $\mathbf{0.42}$ \\
                            & Top-5             & $1.00$ & $1.00$ & $0.70$  & $\mathbf{0.62}$ \\
                            & $\mathcal{K}^{*}$ & $1.16$ & $1.24$ & $27.66$ & $\mathbf{27.89}$ \\ 
    \bottomrule
    \end{tabular}
\end{table}

\Cref{tab:client_defenses_realworld_f1_score} further reports model-agnostic anonymity-set sizes $\mathcal{K}^{*}$ under client defenses (\textbf{\benchkey3}), capturing attacker uncertainty beyond Top-$k$ metrics (higher $\mathcal{K}^{*}$ indicates greater uncertainty).
The estimates show that CL-Tamaraw induces genuine ambiguity rather than merely perturbing the top-ranked guesses: on BBC, despite Top-5$=0.54$, $\mathcal{K}^{*}=26.64$, leaving a substantially broader set of plausible webpages. 
Across \benchkey1--3, CL-Tamaraw is the most robust defense, with FRONT competitive on Amazon and Wikipedia.

\begin{table}[b!]
\caption{Client-side defenses overhead: median (Q1--Q3) of the relative increase ratio $\Delta M$. Baseline averages shown for reference (\benchkey 4).}
\label{tab:client_defenses_overhead}
\centering
    \begin{tabular}{@{}lrrr@{}}
        \toprule
        \textbf{Defense} & 
        \multicolumn{1}{c}{$\Delta\text{Up}$} & 
        \multicolumn{1}{c}{$\Delta\text{Down}$} & 
        \multicolumn{1}{c}{$\Delta T$} \\ 
        \midrule
        HTTPOS       & $0.8\ (0.6 - 1.4)$ & $0.0\ (0.0 - 0.1)$ & $1.0\ (0.3 - 2.1)$ \\
        LLaMA        & $1.9\ (1.7 - 2.6)$ & $2.1\ (1.7 - 2.4)$ & $0.4\ (0.2 - 0.5)$ \\
        FRONT        & $7.4\ (4.4 - 11.1)$ & $5.8\ (3.1 - 8.9)$ & $0.5\ (0.2 - 1.1)$ \\
        CL-TAM       & $6.7\ (2.8 - 11.6)$ & $3.7\ (1.9 - 8.4)$ & $1.0\ (0.5 - 3.0)$ \\
        \midrule
        Baseline avg. &  \multicolumn{1}{c}{$5.37$ KB} &  \multicolumn{1}{c}{$2638.1$ KB } &  \multicolumn{1}{c}{$2.97$ s} \\
        \bottomrule
    \end{tabular}
\end{table}

Finally, \Cref{tab:client_defenses_overhead} summarizes the privacy--cost trade-offs of the client defenses (\textbf{\benchkey4}). 
HTTPOS adds little downstream traffic ($\Delta\text{Down}=0.04$) but increases upload and latency ($\Delta\text{Up}=0.81$, $\Delta T=1.04$), while providing weak privacy. 
LLaMA has the lowest latency overhead ($\Delta T=0.35$) but roughly doubles traffic in both directions and likewise provides limited protection. 
FRONT and CL-Tamaraw incur the largest bandwidth costs: FRONT reaches $\Delta\text{Down}=5.76$, while CL-Tamaraw reaches $3.73$. However, FRONT shows a smaller impact on the perceived latency.
Overall (\benchkey1--4), CL-Tamaraw provides the strongest client-side privacy but at substantial cost, while FRONT offers a lower-latency alternative on Amazon and Wikipedia. More broadly, the dataset-specific defense settings and strongest attackers vary across workloads, underscoring the need for multi-dataset evaluation even when defenses and attackers are individually tuned.

\subsubsection*{\textbf{Takeaways}}
We demonstrated practical emulations of popular client-side defenses using \httptwo primitives, highlighting both their strengths and limitations. 
Client-side defenses uniformly cover all connections within a page load, but cannot directly control download leakage and may incur substantial bandwidth and latency overhead. 
Their effectiveness is also strongly dataset-dependent, requiring defense parameters to be calibrated separately for each dataset rather than using a single operating point. 
Overall, CL-Tamaraw provides the strongest privacy, reaching $\mathcal{K}^{*}=38.43$ on Amazon but dropping to $\mathcal{K}^{*}=2.48$ on Udemy, at median overheads of $\Delta\text{Down}=3.7$ and $\Delta T=1.0$; FRONT provides a lower-latency alternative on Amazon and Wikipedia, with a median $\Delta T=0.5$ but similarly substantial bandwidth overhead.

%% file: sections/4_method_experiments/d_ref_server_defenses.tex
\subsection{\httptwo Server-Side Defenses}
\label{subsection:baseline_fingerprinting_defenses_servers}

In this section, we benchmark established defenses from the perspective of the \httptwo server. The server-side defenses follow the same principles as the client-side defenses: they must be able to pad, split, delay, or add additional noise. We adopt two known fingerprinting defenses, which \emph{can be emulated at the application layer} (1) ALPaCA \cite{DBLP:journals/popets/CherubinHJ17}, a web object morphing strategy; and (2) Tamaraw \cite{DBLP:conf/ccs/CaiNWJG14}, emulated from the server perspective.

At first glance, server-side defenses seem attractive since they could, in principle, protect all users. In practice, however, modern webpages are rarely monolithic. As shown in \Cref{tab:realworld_dataset_summaries} and prior work \cite{DBLP:journals/popets/SibyBWFST23}, most page loads span multiple servers. Unlike clients, which observe and can defend all connections uniformly, servers act independently --- and many may not deploy any protection. 

In our benchmarks, we therefore assume no shared proxy or centralized cloud deployment: each subdomain involved in the webpage load applies defenses individually, while clients follow proactive resource suggestions. The \httptwo adaptations deliver noise via Server Push frames as a proof-of-concept; while some browsers have dropped Server Push support~\cite{wikipediaHTTP2Server}, the mechanism can be replaced by 103 Early Hints --- as demonstrated in \Cref{section:http2_potential}.


\subsubsection{\textbf{ALPaCA}} 
\begin{figure}[t]
        \centering
        \includegraphics[width=\columnwidth]{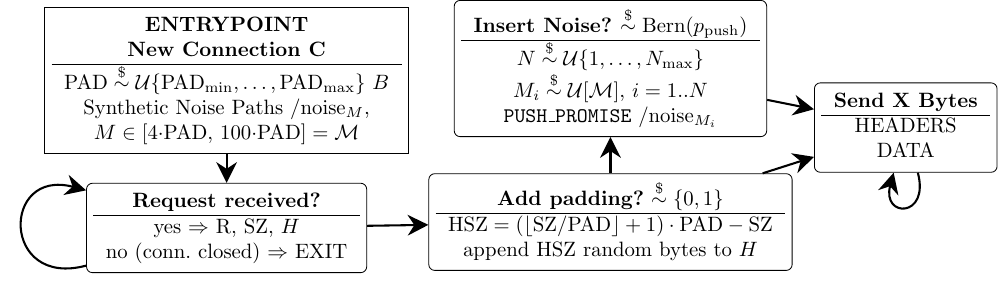}	
        \caption{\httptwo ALPaCA Server-Side Defense \cite{DBLP:journals/popets/CherubinHJ17}.}
        \Description{\httptwo ALPaCA Server-Side Defense \cite{DBLP:journals/popets/CherubinHJ17}.}
        \label{fig:server_side_defense_alpaca}
\end{figure}

The ALPaCA defense \cite{DBLP:journals/popets/CherubinHJ17} supports two modes: a mimicking mode, which reshapes a site toward a larger reference site, and a noise mode, which perturbs resource sizes and injects additional resources. We evaluate the noise mode because it does not require selecting a reference website and maps directly to randomized padding and noise injection in our \httptwo setting.
ALPaCA alters the webpage’s actual content length (images, HTML, CSS, etc.) so that traffic patterns become indistinguishable and unpredictable to a passive adversary. The original design works by (1) padding each object (e.g., an image, CSS file, HTML) with random data and by (2) inserting noise traffic (fake resources) in the main HTML. 
\Cref{fig:server_side_defense_alpaca} illustrates the \httptwo adaptation of the defense. 
Here, object-size perturbation is emulated by injecting dummy bytes into HEADERS frames.
Unlike the original ALPaCA, which embeds fake resource references in HTML, the \httptwo adaptation proactively delivers noisy resources via Server Push frames, without modifying HTML content.

\ifextver
\begin{figure}[b]
    \centering
    \includegraphics[width=\columnwidth]
    {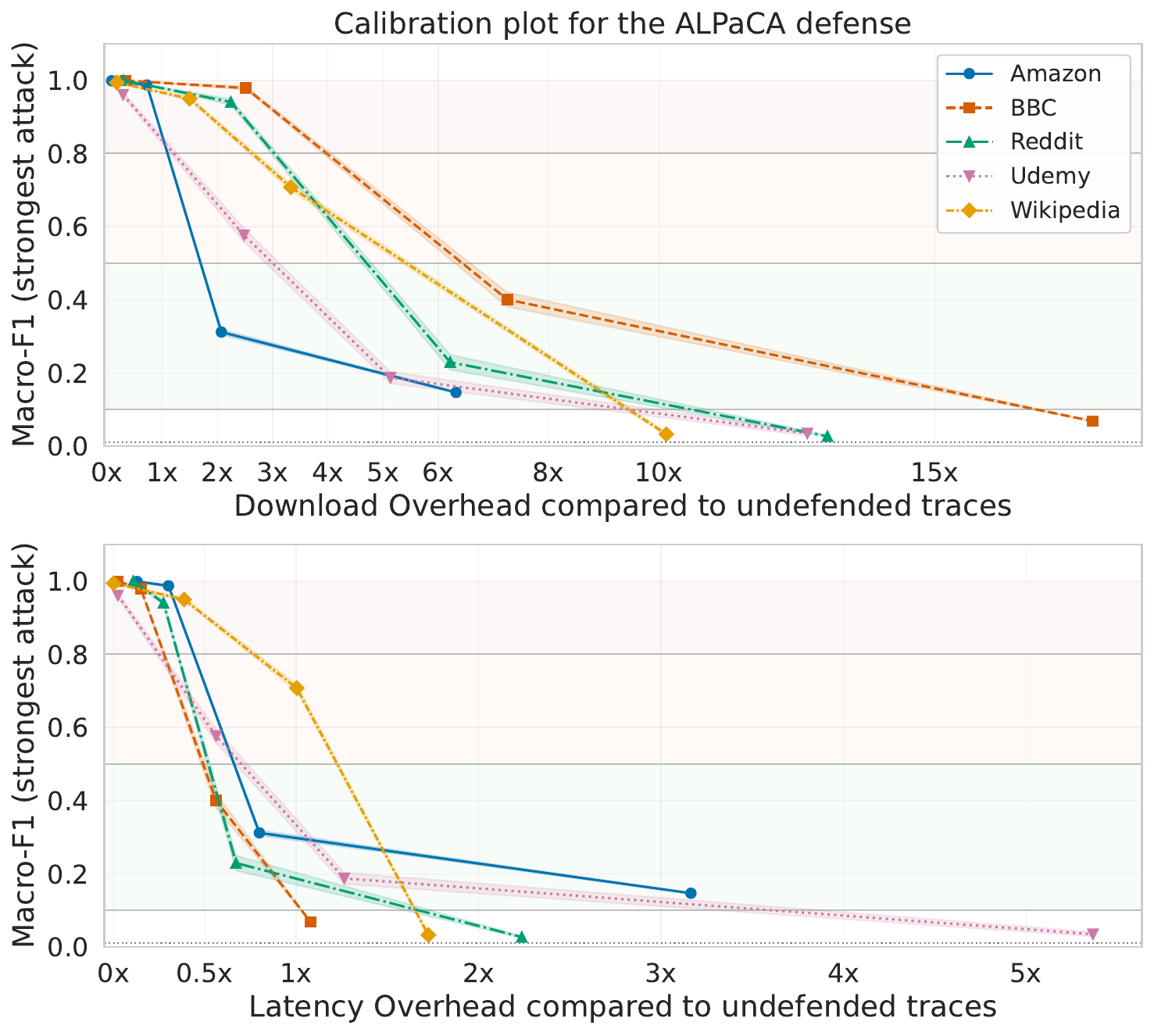}
    \caption{Calibration for the ALPaCA defense (\benchkey0).
    Strongest-attacker Macro-F1 per dataset against bandwidth (top) and latency (bottom) overhead. Horizontal lines mark the weak, moderate, and strong attacker thresholds; the dotted line is random guessing.}
    \Description{Calibration for the ALPaCA defense (\benchkey0).}
    \label{fig:server_defense_calibration_alpaca}
\end{figure}
\fi

For the ALPaCA defense calibration (\textbf{\benchkey0}), we jointly vary ALPaCA's per-connection padding range, maximum number of pushed noise objects, and push probability $p_{\mathrm{push}}$.
Calibration is performed with ALPaCA enabled on all connections involved in the page load.
The three parameters scale together because each strengthens the same padding-and-noise mechanism, and noise-object sizes are defined relative to the padding granularity.
\ifextver
The four candidate configurations use padding ranges $[256,1024]$, $[512,4000]$, $[768,6000]$, and $[1024,8000]$\,B, with corresponding maximum numbers of noise objects $2$, $4$, $6$, and $10$, and $p_{\mathrm{push}}\in\{0.2,0.3,0.4,0.5\}$, respectively.
\Cref{fig:server_defense_calibration_alpaca} shows that ALPaCA requires relatively aggressive padding before providing substantial protection.
Increasing the padding range from $[768,6000]$\,B to $[1024,8000]$\,B, the maximum number of noise objects from $6$ to $10$, and $p_{\mathrm{push}}$ from $0.4$ to $0.5$ reduces the calibration Macro-F1 from $0.312$ to $0.147$ on Amazon, $0.401$ to $0.068$ on BBC,
$0.229$ to $0.027$ on Reddit, $0.187$ to $0.035$ on Udemy, and $0.708$ to $0.033$ on Wikipedia.
Although this increase also incurs substantial downstream and latency overhead, the additional privacy gain remains large on every dataset.
\fi
The calibration shows a sharp transition: padding drawn from $[1024,8000]$\,B, with up to $10$ noise objects and $p_{\mathrm{push}}=0.5$, reduces calibration Macro-F1 below $0.15$ on every dataset, whereas the immediately weaker configuration remains
between $0.19$ and $0.71$. We therefore select this defense configuration for all five datasets.

\begin{figure}[h]
        \centering
        \includegraphics[width=\columnwidth]{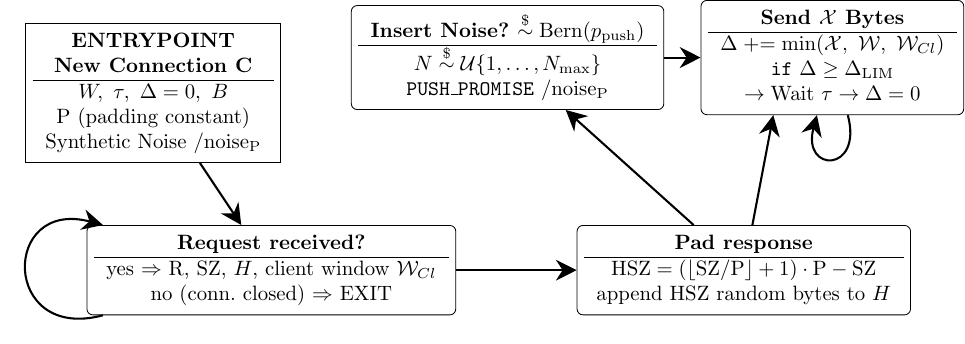}	
        \caption{\httptwo emulation of the SRV-Tamaraw Defense \cite{DBLP:conf/ccs/CaiNWJG14}.  }
        \Description{\httptwo emulation of the SRV-Tamaraw Defense \cite{DBLP:conf/ccs/CaiNWJG14}.  }
        \label{fig:server_side_defense_tamaraw}
\end{figure}
\ifextver
\begin{figure}[b]
    \centering
    \includegraphics[width=\columnwidth]
    {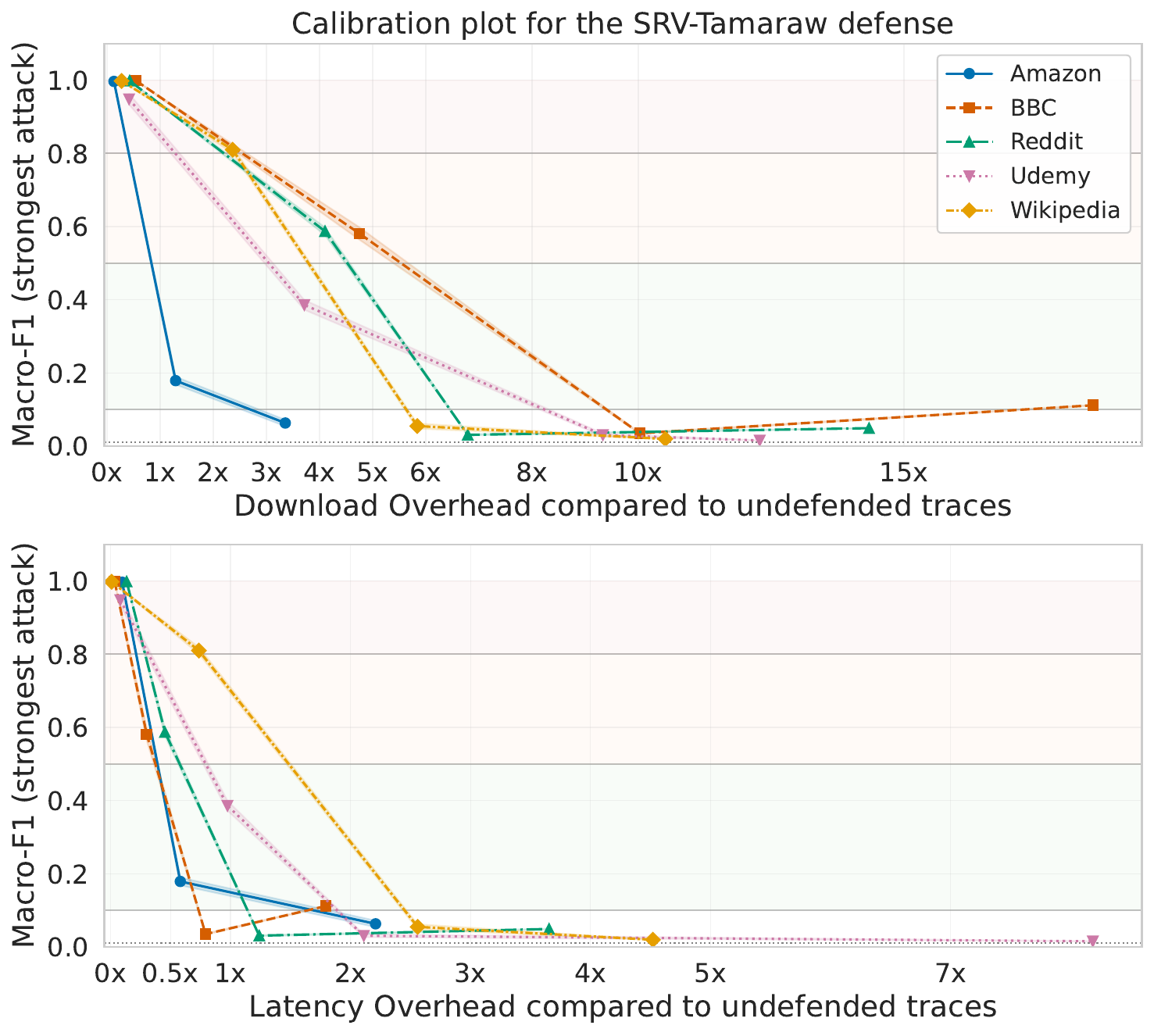}
    \caption{Calibration for the SRV-Tamaraw defense (\benchkey0).
    Strongest-attacker Macro-F1 per dataset against bandwidth (top) and latency (bottom) overhead. Horizontal lines mark the weak, moderate, and strong attacker thresholds; the dotted line is random guessing.}
    \Description{Calibration for the SRV-Tamaraw defense (\benchkey0).}
    \label{fig:server_defense_calibration_tamaraw}
\end{figure}
\fi
\subsubsection{\textbf{SRV-Tamaraw (SRV-TAM)}}
We emulate Tamaraw from the server side using three mechanisms: (1) noise injection via Server Push, (2) response padding to a fixed-size multiple, and (3) paced transmission.
Unlike ALPaCA's randomized per-connection padding, SRV-Tamaraw uses a fixed padding size to enforce more uniform traffic shaping.
The server tracks transmitted bytes with a counter $\Delta$ and pauses for $\tau$ whenever $\Delta$ reaches the threshold $B$, before resetting the counter.
\Cref{fig:server_side_defense_tamaraw} illustrates the \httptwo adaptation.

For the SRV-Tamaraw defense calibration (\textbf{\benchkey0}), we jointly vary the padding constant $P$, output flow-control window $W$, send delay $\tau$, pacing threshold $B$, maximum number of pushed noise objects $N_{\max}$, and push probability $p_{\mathrm{push}}$.
These parameters scale together to increase shaping intensity through larger padding, tighter flow control, longer pauses, and more noise.
\ifextver
The four candidate configurations use $(P,W,\tau,B,N_{\max},p_{\mathrm{push}})$ values
$(1024,\allowbreak 16384,\allowbreak 0.2,\allowbreak 16384,\allowbreak 2,\allowbreak 0.2)$,
$(4096,\allowbreak 8192,\allowbreak 0.5,\allowbreak 8192,\allowbreak 4,\allowbreak 0.3)$,
$(6144,\allowbreak 4096,\allowbreak 0.8,\allowbreak 6144,\allowbreak 6,\allowbreak 0.4)$, and
$(8092,\allowbreak 2048,\allowbreak 1.0,\allowbreak 4096,\allowbreak 10,\allowbreak 0.5)$, respectively, where $P$, $W$, and $B$ are in bytes and $\tau$ is in milliseconds.
\Cref{fig:server_defense_calibration_tamaraw} shows a clear privacy--overhead knee at $P=6144$\,B, $W=4096$\,B, $\tau=0.8$\,ms, $B=6144$\,B, $N_{\max}=6$, and $p_{\mathrm{push}}=0.4$.
On Amazon, this configuration reaches a calibration Macro-F1 of $0.063$.
On BBC and Reddit, it reaches $0.035$ and $0.031$, respectively, and the stronger configuration does not improve protection.
On Udemy, the stronger configuration reduces Macro-F1 only from $0.031$ to $0.015$, while latency overhead increases from $2.11$ to $8.19$.
Similarly, on Wikipedia, it reduces Macro-F1 from $0.055$ to $0.020$ while downstream overhead increases from $5.84$ to $10.51$ and latency overhead from $2.56$ to $4.52$.

\fi
The calibration shows a clear knee at $P=6144$\,B, $W=4096$\,B, $\tau=0.8$\,ms, $B=6144$\,B, $N_{\max}=6$, and $p_{\mathrm{push}}=0.4$, reducing calibration Macro-F1 to $0.031$--$0.063$ across all five datasets.
Stronger shaping provides limited additional protection at substantially higher overhead, so we select this configuration for all datasets.

\subsubsection{\textbf{Fingerprinting on Server-Defended Datasets}}

For each website, we benchmark the defenses applied to either the main page connection ($1^\text{st}$ Party), the leakiest third-party server ($\text{CDN}$), or all servers involved (All).

\begin{table}[t]
\caption{Baseline resilience of the \httptwo \emph{Server Defenses} against the best-performing evaluated attacker (\benchkey1 and 2), together with the estimator-derived anonymity-set proxy $\mathcal{K}^{*}$ (\benchkey3). The best scores among single-server deployments are underlined, while the best-performing deployments overall are highlighted in bold.
Macro-F1 and Top-5 are means; their 95\% CIs are below $0.02$ and omitted.}
\label{tab:server_defenses_realworld_f1_score}
    \begin{tabular}{@{}c@{\,}c@{\,}ccc@{\,}|@{\,}ccc@{}} \toprule
    \multirow{2}{*}{\textbf{Data}} & \multirow{2}{*}{\textbf{Metric}} & \multicolumn{3}{c@{\,}|@{\,}}{\textbf{ALPaCA}}     & \multicolumn{3}{@{\,}c}{\textbf{SRV-TAM}}     \\
                                      &                   & $1^\text{st}$ Party & $\text{CDN}$ & All & $1^\text{st}$ Party & $\text{CDN}$ & All \\ \midrule
    \multirow{3}{*}{Amz.}
                                      & F1                & $1.00$ & $0.76$ & $0.21$ & $1.00$ & $\underline{0.26}$ & $\mathbf{0.08}$ \\
                                      & Top-5             & $1.00$ & $0.94$ & $0.59$ & $1.00$ & $\underline{0.51}$ & $\mathbf{0.30}$ \\
                                      & $\mathcal{K}^{*}$ & $1.01$ & $5.21$ & $62.15$ & $1.02$ & $\underline{24.47}$ & $\mathbf{77.10}$ \\
    \midrule
    \multirow{3}{*}{BBC}
                                      & F1                & $0.99$ & $\underline{0.87}$ & $0.20$ & $0.94$ & $0.90$ & $\mathbf{0.15}$ \\
                                      & Top-5             & $1.00$ & $0.99$ & $0.54$ & $\underline{0.96}$ & $0.98$ & $\mathbf{0.43}$ \\
                                      & $\mathcal{K}^{*}$ & $1.06$ & $\underline{2.32}$ & $23.56$ & $1.62$ & $1.70$ & $\mathbf{63.08}$ \\
    \midrule
    \multirow{3}{*}{Reddit}
                                      & F1                & $\underline{0.33}$ & $0.97$ & $\mathbf{0.04}$ & $0.76$ & $0.98$ & $0.06$ \\
                                      & Top-5             & $\underline{0.45}$ & $1.00$ & $\mathbf{0.18}$ & $0.92$ & $1.00$ & $0.22$ \\
                                      & $\mathcal{K}^{*}$ & $\underline{39.31}$ & $1.37$ & $\mathbf{83.42}$ & $5.23$ & $1.23$ & $82.19$ \\
    \midrule
    \multirow{3}{*}{Udemy}
                                      & F1                & $0.64$ & $0.49$ & $0.08$ & $\underline{0.41}$ & $0.83$ & $\mathbf{0.03}$ \\
                                      & Top-5             & $0.91$ & $0.85$ & $0.27$ & $\underline{0.54}$ & $0.99$ & $\mathbf{0.12}$ \\
                                      & $\mathcal{K}^{*}$ & $8.14$ & $9.46$ & $57.93$ & $\underline{14.55}$ & $3.43$ & $\mathbf{97.51}$ \\
    \midrule
    \multirow{3}{*}{Wiki}
                                      & F1                & $\underline{0.09}$ & $0.97$ & $\mathbf{0.06}$ & $0.69$ & $0.97$ & $0.08$ \\
                                      & Top-5             & $\mathbf{0.17}$ & $1.00$ & $0.20$ & $0.93$ & $0.99$ & $0.25$ \\
                                      & $\mathcal{K}^{*}$ & $\mathbf{76.26}$ & $1.36$ & $65.72$ & $7.83$ & $1.34$ & $74.92$ \\
    \bottomrule
    \end{tabular}
\end{table}

\Cref{tab:server_defenses_realworld_f1_score} reports the privacy of the calibrated server-side defenses against the strongest hyperparameter-tuned attacker for each dataset--defense pair (\textbf{\benchkey1} and \textbf{\benchkey2}).
Placement remains the dominant factor.
On Amazon, defending the CDN connection is already highly effective for SRV-TAM (F1 $=0.26$, $\mathcal{K}^{*}=24.47$), whereas defending only the first-party server provides essentially no protection.
BBC shows the opposite extreme: neither single-server placement is sufficient, but deployment across all servers reduces F1 to $0.20$ for ALPaCA and $0.15$ for SRV-TAM.
Reddit, Udemy, and Wikipedia exhibit stronger first-party leakage: ALPaCA reaches F1 $=0.33$ and $0.09$ on Reddit and Wikipedia, respectively, while SRV-TAM reaches $0.41$ on Udemy.

The anonymity-set estimates (\textbf{\benchkey3}) reinforce this placement effect.
Targeting a single dominant connection can already provide substantial ambiguity---for example, $\mathcal{K}^{*}=24.47$ for SRV-TAM on Amazon's CDN, $39.31$ for ALPaCA on Reddit's first-party server, and $76.26$ for ALPaCA on Wikipedia's first-party server.
When leakage is distributed across connections, however, full deployment is necessary: on BBC, $\mathcal{K}^{*}$ rises from at most $2.32$ under any single-server placement to $23.56$ with ALPaCA-All and $63.08$ with SRV-TAM-All.
Full deployment is also particularly effective on Udemy, where SRV-TAM reaches F1 $=0.03$ and $\mathcal{K}^{*}=97.51$.
Overall, the results show that server-side defenses can provide strong privacy, but their effectiveness depends critically on placing the defense at the connections carrying the dominant fingerprinting signal rather than simply instrumenting the first-party server.

\begin{table}[b!]
\caption{Server-side defenses overhead ratio: median (Q1--Q3) of the relative increase ratio $\Delta M$ pooled across all five case studies, for single-server (top) and all-servers (bottom) deployments. (\benchkey4). $\Delta\text{Up} = 0$ throughout, as the client sends no additional requests.}  
\label{tab:server_defenses_overhead}
\centering
    \begin{tabular}{@{}llrr@{}}
        \toprule
        \multicolumn{2}{c}{\textbf{Defense}} & 
        \multicolumn{1}{c}{$\Delta\text{Down}$} & 
        \multicolumn{1}{c}{$\Delta T$} \\ 
        \midrule
        \multirow{2}{*}{ALPaCA}  
            & Single Srv. & $4.9\ (2.9 - 7.8)$   &  $0.7\ (0.4 - 1.7)$ \\
            & All Srv.    & $11.7\ (7.1 - 15.8)$ &  $2.4\ (1.4 - 4.2)$ \\
        \midrule
        \multirow{2}{*}{SRV-TAM} 
            & Single Srv. & $3.6\ (2.1 - 6.3)$   & $0.5\ (0.3 - 1.5)$ \\
            & All Srv.    & $6.8\ (4.4 - 9.6)$   & $1.7\ (1.0 - 2.6)$ \\
        \midrule
        \multicolumn{2}{c}{Baseline avg.} &  \multicolumn{1}{c}{$2638.10$ KB} &  \multicolumn{1}{c}{$2.97$ s} \\
        \bottomrule
    \end{tabular}
\end{table}

\Cref{tab:server_defenses_overhead} reports the overhead of the calibrated server-side defenses (\textbf{\benchkey4}).
Both defenses incur zero upload overhead because the client sends no additional requests.
Because each defense is calibrated per dataset toward its strongest practical privacy operating point, the reported overhead also reflects the cost of the selected protection level rather than a matched-cost comparison.
In these conditions, SRV-TAM is nevertheless consistently cheaper than ALPaCA.
Thus, ALPaCA's higher overhead partly reflects the stronger configuration selected during calibration, while SRV-TAM provides the more favorable privacy--overhead trade-off overall.

\subsubsection{\textbf{Takeaways}}
We demonstrated practical emulations of established server-side defenses using \httptwo features, with defense parameters calibrated per dataset and attackers' hyperparameters tuned independently for each dataset--defense pair.
The results remain strongly dataset- and placement-dependent: the most effective single-server deployment varies across case studies, while protecting all participating servers generally provides the strongest privacy at higher cost.
Overall, SRV-TAM provides the stronger privacy--overhead trade-off for full deployment, while targeted single-server defenses can be highly effective when one connection dominates the leakage (e.g., Amazon, Reddit, Udemy or Wikipedia).

%% file: sections/4_method_experiments/e_http2_opportunities_analysis.tex
\section{The Untapped Potential of \httptwo for Fingerprinting Defenses}
\label{section:http2_potential}

We showed that \httptwo users can emulate \emph{established} WF defenses, achieving privacy improvements across all case studies and deployment perspectives (client and server). Yet, these guarantees often come at considerable overhead or depend on carefully targeting the right servers --- limitations that stem from the fact that these defenses were never designed with \httptwo’s architecture in mind.

\subsection{Opportunities with \httptwo}
\label{subsection:http2_new_opprotunities}

\noindent\emph{Can we do better?} In this section, we outline various novel strategies for leveraging \httptwo features to raise the baseline privacy of page loads and to overcome key limitations of existing defense designs.


\subsubsection{\textbf{Client Side}} Most defenses assume noise is essential, overlooking the privacy benefits inherent in application-layer behavior, such as multiplexing and flow control. 
We first detail \httptwo's features, which can improve privacy from the client perspective, then discuss how to insert noise more efficiently.

\begin{fallacybox}
\textbf{Client Opportunity 1:} Randomizing client behavior by using \httptwo\ features can improve the baseline privacy.
\end{fallacybox}

\noindent A light privacy-conscious \httptwo client (H2PC) can leverage the following features:
\begin{itemize}[left=0pt,noitemsep]
    \item \emph{Multiplexing \& Prioritization.} The client can shuffle and batch a subset of the pending requests, proactively altering the burst patterns of the connection. 
    \item \emph{Flow Control.} While the Tamaraw defense uses flow control for uniform traffic shaping, we can also turn it into an unpredictability-driven defense. Concretely, the client can randomize flow control window sizes per connection, enforcing different burst patterns on each page reload.
    \item \emph{Connection Probing.} \texttt{PING} frames carry only 8 bytes, so they minimally affect bandwidth. 
    As long as the mechanism is not abused, inserting a random number of PINGs disrupts burst patterns when combined with prior techniques.
\end{itemize}

\begin{fallacybox}
\textbf{Client Opportunity 2:} \httptwo\ primitives allow clients to create `guarding' noise streams.
\end{fallacybox}

A key lesson from existing defenses is that more noise generally yields stronger privacy. 
For example, if we vary the noise quantity ceiling in the FRONT defense for the Udemy dataset between $10 \rightarrow 500$, we get a monotonic F1 score variation between $[0.48 - 0.76]$.
Yet, the critical question of ``how much noise is enough'' remains largely unaddressed. 
To move beyond brute-force noise, we leverage \httptwo features against primary leakage sources, yielding more efficient and scalable defenses.

\begin{figure}[t!]
        \centering
        \includegraphics[width=\columnwidth]{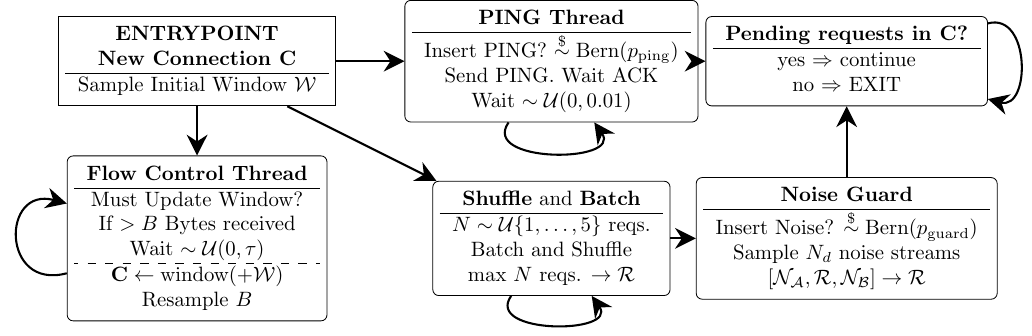}	
        \caption{A privacy-conscious \httptwo client (H2PC) flow.  }
        \Description{A privacy-conscious \httptwo client (H2PC) flow.  }
        \label{fig:client_side_http2_mods}
\end{figure}

A notable idea comes from the HTTPOS \cite{DBLP:conf/ndss/LuoZCLCP11} defense, which focuses on binary resources --- often the leakiest. Its strategy can be generalized: large resources can be guarded by surrounding them with noise streams, allowing \httptwo to multiplex guarding streams alongside legitimate traffic, with the frame scheduler interleaving DATA frames from multiple streams in a pattern that depends on window sizes, priorities, and implementation-specific scheduling. 
Unlike HTTPOS, this approach is not limited to binary content and does not require server cooperation. And unlike LLaMA, FRONT, or CL-Tamaraw, it targets the sensitive streams directly rather than relying on opportunistic overlaps between noise and real traffic. 

\Cref{fig:client_side_http2_mods} depicts the client-side defense workflow.
For each connection \textbf{C}, the client (i) randomizes the initial receive window $\mathcal{W}$ using the flow-control parameters and (ii) launches an independent PING thread whose frames, when serialized into \tcp segments, probabilistically interleave with DATA-frame segments, perturbing observable burst-direction patterns.
Pending requests are then grouped into randomized batches and reordered (as in LLaMA), while each request may be accompanied by sampled guarding noise, altering CUMUL and packet-level statistics.
After receiving $B$ bytes, the client introduces additional randomness by delaying the receive-window update by a value sampled from $\mathcal{U}(0,\tau)$ and resampling $B$, further obscuring timing patterns.
Resampling prevents window updates from occurring at deterministic byte intervals, which could otherwise become a learnable signature.

For defense calibration (\textbf{\benchkey0}), we calibrate five parameters: the guarding-stream limit $N_d$, PING probability $p_{\mathrm{ping}}$ and count range, receive-delay bound $\tau$, and receive-delay threshold $B$.
The two weakest configurations use no guarding streams, isolating the effect of \httptwo-native randomization before guarding noise is introduced: the first uses randomized flow control alone, while the second adds PING padding, both with negligible measured bandwidth overhead.
%
\ifextver

The six candidate configurations progressively increase the guarding-stream and PING activity while tightening the receive-delay parameters.
The first uses $N_d=0$, $p_{\mathrm{ping}}=0$, PING count $[1,1]$, $\tau=50\,\mu$s, and $B=20000$\,B; 
the second uses $N_d=0$ $p_{\mathrm{ping}}=0.25$, PING count $[1,2]$, $\tau=50\,\mu$s, and $B=20000$\,B; 
and the third uses $N_d=1$, $p_{\mathrm{ping}}=0.35$, PING count $[1,2]$, $\tau=80\,\mu$s, and $B=15000$\,B.
The remaining configurations use $(N_d,p_{\mathrm{ping}})=(1,0.50)$, $(2,0.70)$, and $(3,1.00)$, with PING-count ranges $[1,3]$, $[1,5]$, and $[2,8]$, receive-delay bounds of $100$, $200$, and $500\,\mu$s, and thresholds of $10000$, $5000$, and $2500$\,B, respectively.
\begin{figure}[t]
    \centering
    \includegraphics[width=\columnwidth]
    {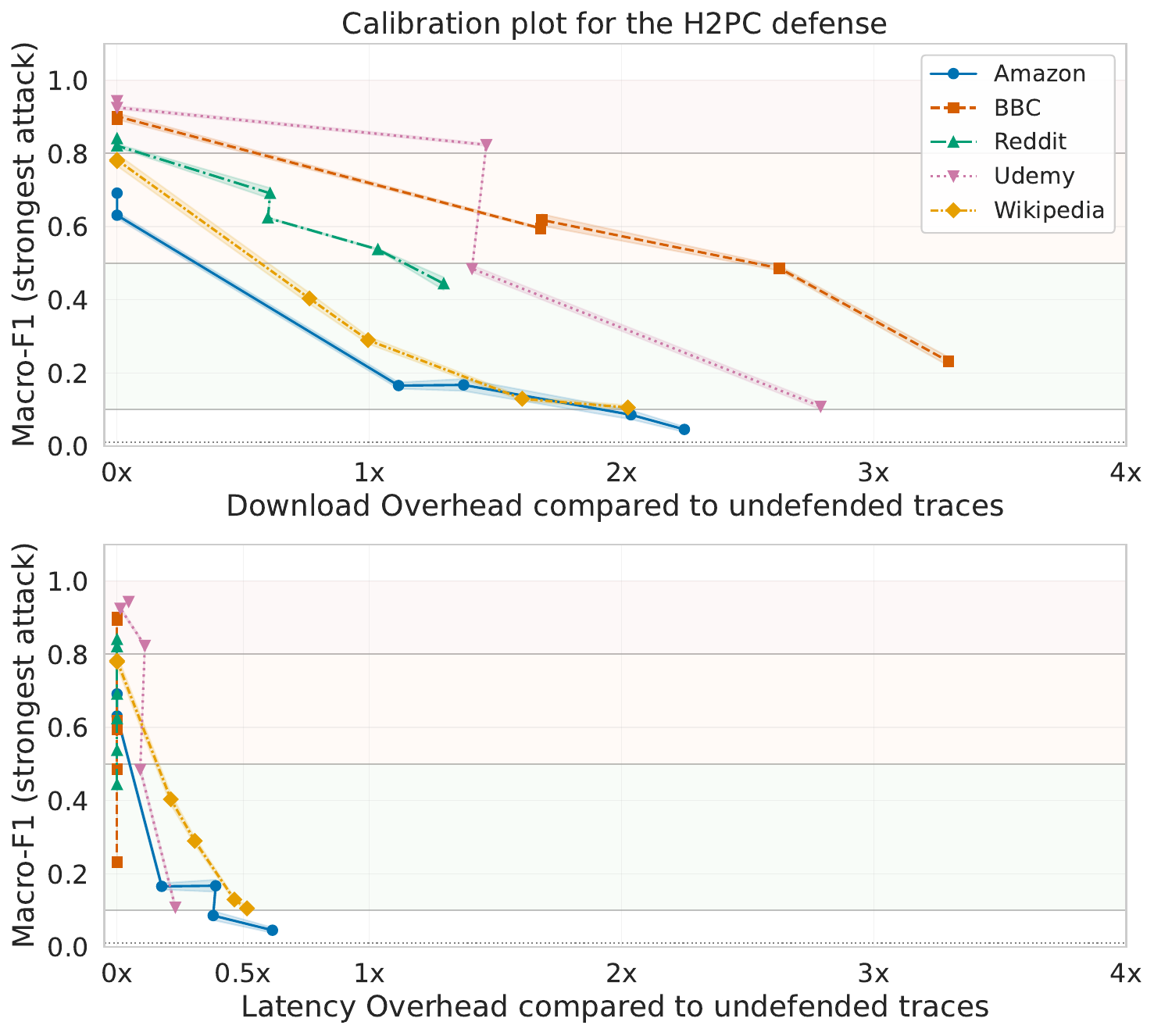}
    \caption{Calibration for the H2PC defense (\benchkey0).
    Strongest-attacker Macro-F1 per dataset against bandwidth (top) and latency (bottom) overhead. Horizontal lines mark the weak, moderate, and strong attacker thresholds; the dotted line is random guessing.}
    \Description{Calibration for the H2PC defense (\benchkey0).}
    \label{fig:client_defense_calibration_h2pc}
\end{figure}
\Cref{fig:client_defense_calibration_h2pc} shows a comparatively consistent reduction in attacker performance as H2PC intensity increases.
On Amazon, the calibration Macro-F1 decreases from $0.167$ with $N_d=1$, $p_{\mathrm{ping}}=0.5$, $\tau=100\,\mu$s, and $B=10000$\,B to $0.086$ with $N_d=2$, $p_{\mathrm{ping}}=0.7$, $\tau=200\,\mu$s, and $B=5000$\,B; the strongest configuration further reduces it to $0.046$, but with additional bandwidth and latency overhead.
Wikipedia shows a similar knee, decreasing from $0.290$ to $0.130$ and then $0.105$ across the same configurations.
In contrast, BBC, Reddit, and Udemy continue to obtain substantial privacy gains at the strongest configuration, reaching Macro-F1 values of $0.232$, $0.444$, and $0.109$, respectively.

\fi
Calibration reveals two operating regimes: Amazon and Wikipedia favor an intermediate privacy--overhead point, whereas BBC, Reddit, and Udemy justify the strongest configuration.
For Amazon and Wikipedia, we select $N_d=2$, $p_{\mathrm{ping}}=0.7$, PING count $[1,5]$, $\tau=200\,\mu$s, and $B=5000$\,B, reaching calibration Macro-F1 $\leq 0.13$.
For BBC, Reddit, and Udemy, we select the strongest configuration, $N_d=3$, $p_{\mathrm{ping}}=1.0$, PING count $[2,8]$, $\tau=500\,\mu$s, and $B=2500$\,B, reaching calibration Macro-F1 $\leq 0.44$.

\begin{table}[t!]
\caption{Client-Side Defense Landscape with \httptwo. H2PC vs. best performing defense (\benchkey1, 2, 3, 4). Macro-F1 and Top-5 are means; their 95\% CIs are below 0.02 and omitted.}
\label{tab:client_mod_realworld_anon_sets}
\centering
\resizebox{\columnwidth}{!}{
    \begin{tabular}{@{}ccrrr|rrr@{}}
    \toprule
    \textbf{Dataset}        & \textbf{Defense}  & \multicolumn{1}{c}{\textbf{F1}}           & \multicolumn{1}{c}{\textbf{Top-5}}  & \multicolumn{1}{c}{$\boldsymbol{\mathcal{K}^{*}}$} &   \multicolumn{1}{c}{$\mathbf{\Delta \textbf{Up}}$} &   \multicolumn{1}{c}{$\mathbf{\Delta \textbf{Down}}$} &  \multicolumn{1}{c}{$\mathbf{\Delta T}$}  \\ \midrule
    \multirow{2}{*}{Amz.}
                             & CL-TAM   & $0.24$ & $0.51$ & $38.4$     & $3.3$     & $3.2$     & $1.4$ \\
                             & H2PC     & $0.32$ & $0.60$ & $20.1$     & $2.1$     & $2.0$     & $0.4$ \\
    \midrule
    \multirow{2}{*}{BBC}
                             & CL-TAM   & $0.24$ & $0.54$ & $26.6$     & $2.5$     & $4.1$     & $0.3$ \\
                             & H2PC     & $0.51$ & $0.77$ & $10.8$     & $2.3$     & $3.3$     & $0.01$ \\
    \midrule
    \multirow{2}{*}{Reddit}
                             & CL-TAM   & $0.38$ & $0.47$ & $34.1$     & $16.3$    & $3.0$     & $0.6$ \\
                             & H2PC     & $0.68$ & $0.84$ & $8.11$     & $1.1$     & $1.3$     & $0.01$ \\
    \midrule
    \multirow{2}{*}{Udemy}
                             & CL-TAM   & $0.88$ & $0.99$ & $2.48$     & $10.5$    & $8.2$     & $4.7$ \\
                             & H2PC     & $0.70$ & $0.87$ & $7.31$     & $2.8$     & $2.8$     & $0.2$ \\
    \midrule
    \multirow{2}{*}{Wiki}
                             & CL-TAM   & $0.42$ & $0.62$ & $27.9$     & $6.6$     & $2.2$     & $1.1$ \\
                             & H2PC     & $0.70$ & $0.87$ & $7.32$     & $1.4$     & $1.6$     & $0.5$ \\  \bottomrule

    \end{tabular}
}
\end{table}

\Cref{tab:client_mod_realworld_anon_sets} compares H2PC against CL-Tamaraw, the strongest client-side defense overall, across privacy and overhead (\textbf{\benchkey1--4}).
H2PC keeps attacker Macro-F1 at or below $0.70$ on all five datasets while substantially reducing latency and downstream overhead.
On Amazon, BBC, Reddit, and Wikipedia, this yields a lower-cost privacy operating point: H2PC maintains $\mathcal{K}^{*}\geq 7.32$ while reducing $\Delta T$ and $\Delta\text{Down}$ relative to CL-Tamaraw.
The trade-off is especially favorable on Udemy, where H2PC improves $\mathcal{K}^{*}$ from $2.48$ to $7.31$, while reducing $\Delta T$ from $4.7$ to $0.2$ and $\Delta\text{Down}$ from $8.2$ to $2.8$.
Overall, H2PC provides competitive privacy at substantially lower cost.


\subsubsection{\textbf{Server Side}} We conduct a similar analysis on the privacy-preserving impact of \httptwo features from the server perspective.

\begin{fallacybox}
\textbf{Server Opportunity 1:} Randomizing \httptwo's built-in feature behavior can improve web-browsing privacy.
\end{fallacybox}

\noindent A privacy-conscious \httptwo server (H2PS) can use:

\begin{itemize}[left=0pt,noitemsep]
    \item \emph{Multiplexing \& Prioritization.} The server can buffer multiple requests or delay streams, leading to unpredictable burst patterns.
    \item \emph{Flow Control.} \httptwo servers can artificially constrain their transmission rate by maintaining an internal flow-control limit smaller than the receiver’s advertised window, effectively under-utilizing the available window capacity.
    \item \emph{Connection Probing.} Similar to clients, servers can send \texttt{PING} frames to disrupt burst patterns.
\end{itemize}

\begin{fallacybox}
\textbf{Server Opportunity 2:} \httptwo\ primitives allow $1^\text{st}$ Party servers to defend the entire webpage.
\end{fallacybox}

When a \httptwo-conformant client prefetches suggested resources, servers can suggest noise traffic using mechanisms such as Server-Push or ``103 Early Hints'' (as with ALPaCA and SRV-Tamaraw). 
 Notably, ``103 Early Hints'' \cite{DBLP:journals/rfc/rfc8297} work across servers: a server can reference resources from any domain, which a client would fetch. This adds a new dimension to server-side defenses, enabling first parties to protect the entire webpage. 
 
 \begin{figure}[b]
        \centering
        \includegraphics[width=\columnwidth]{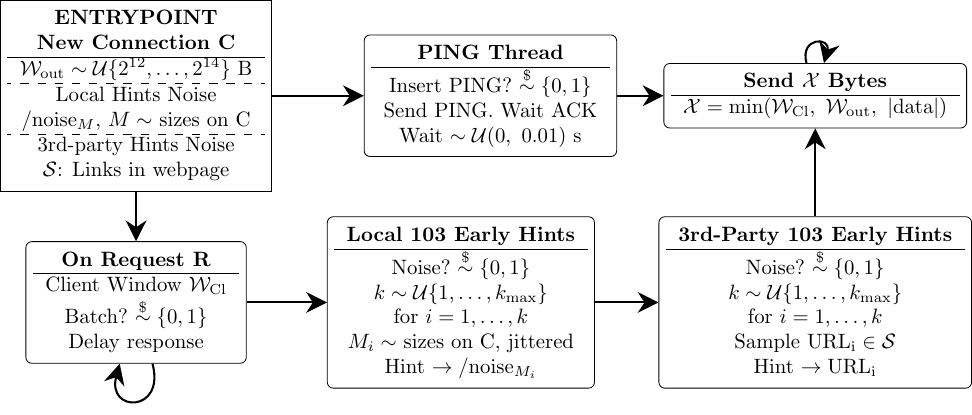}	
        \caption{A privacy-conscious \httptwo server (H2PS) flow.}
        \Description{A privacy-conscious \httptwo server (H2PS) flow.}
        \label{fig:server_side_http2_mods}
\end{figure}

\Cref{fig:server_side_http2_mods} shows the H2PS defense workflow. 
For each new connection $C$ to the $1^{st}-$Party (e.g., www.bbc.com), the server randomizes its outbound flow-control limit $\mathcal{W}_{\mathrm{out}}$ and spawns an independent PING thread to disrupt burst patterns, similar to H2PC.
For every incoming request, the server probabilistically issues two types of ``103 Early Hints'' before returning the actual response: (1) synthetic local hints, served directly by the defended $1^{\text{st}}$ server; and (2) third-party hints, with real resources hosted on other CDN servers observed in the page load (e.g., by extracting them from the HTML).
This is a key advantage of ``Early Hints'' over ``Server Push'': as Early Hints send only \texttt{Link} headers rather than actual content, the $1^{\text{st}}$-party server can hint resources on any origin without CDN cooperation or a shared proxy. Assuming client cooperation, the hint fetches are multiplexed with legitimate \httptwo frames, perturbing the network metadata observed by a passive adversary. 
Requests are also probabilistically batched and delayed, leading to timing unpredictability. 

For defense calibration (\textbf{\benchkey0}), H2PS varies the number of ``103 Early Hints'' per defended connection ($[1,1]$ to $[40,120]$), PING padding, and HPACK randomization at the strongest configurations.
\ifextver
The eight candidate configurations use Early Hints ranges $[1,1]$, $[1,2]$, $[1,5]$, $[1,10]$, $[6,18]$, $[12,36]$, $[24,72]$, and $[40,120]$.
The first configuration uses no PING padding; all remaining configurations use one PING, and HPACK randomization is enabled for the three strongest ranges, starting at $[12,36]$.
\begin{figure}[t]
    \centering
    \includegraphics[width=\columnwidth]
    {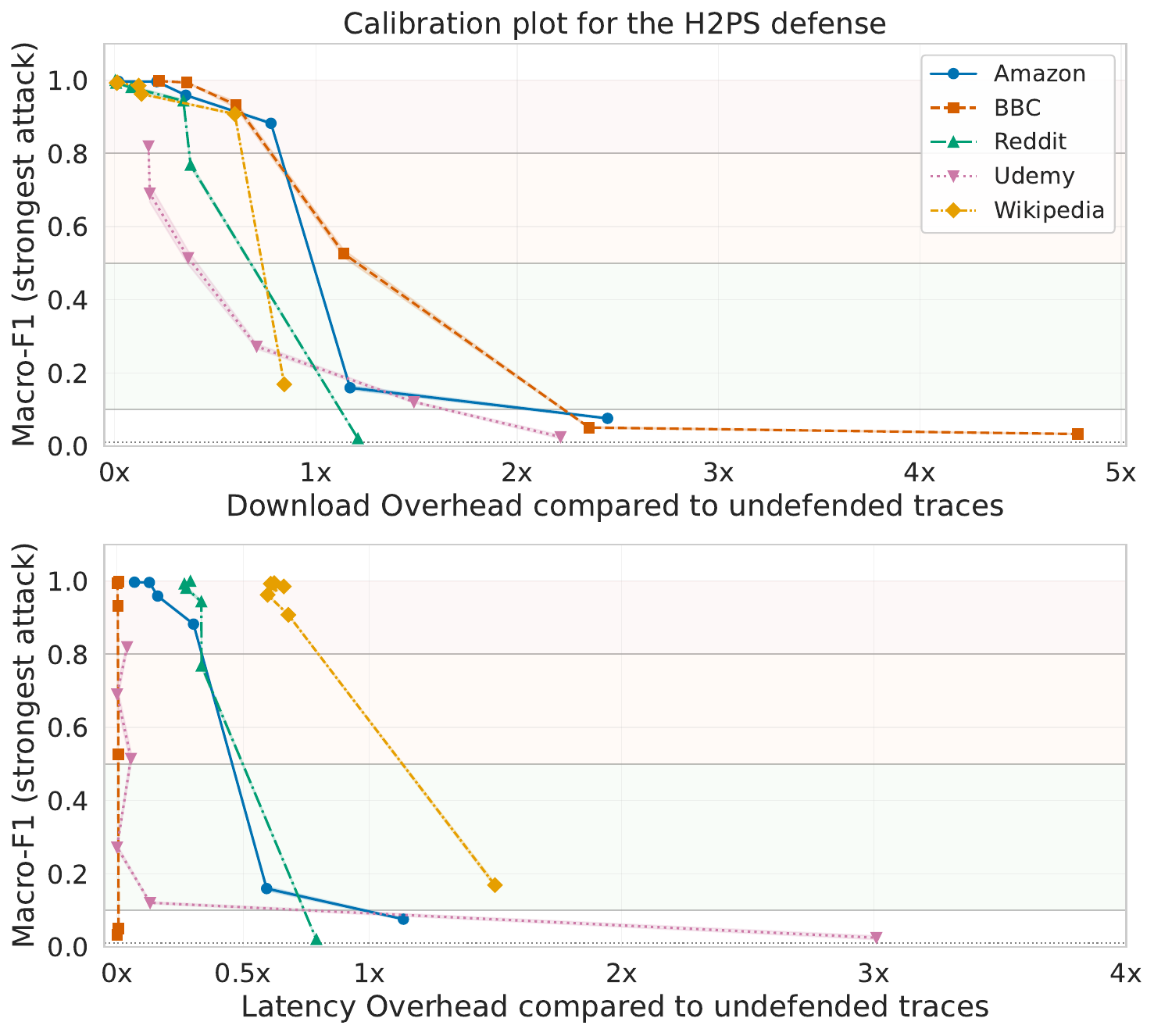}
    \caption{Calibration for the H2PS defense (\benchkey0).
    Strongest-attacker Macro-F1 per dataset against bandwidth (top) and latency (bottom) overhead. Horizontal lines mark the weak, moderate, and strong attacker thresholds; the dotted line is random guessing.}
    \Description{Calibration for the H2PS defense (\benchkey0).}
    \label{fig:server_defense_calibration_h2ps}
\end{figure}
\Cref{fig:server_defense_calibration_h2ps} shows that lighter H2PS configurations provide limited protection, with dataset-dependent knees at stronger configurations.
Amazon and Reddit select the $[12,36]$ Early Hints range, reaching calibration Macro-F1 of $0.076$ and $0.021$, respectively.
BBC and Udemy select $[6,18]$, reaching $0.051$ and $0.121$, as stronger configurations incur substantially higher overhead for limited additional benefit.
Wikipedia requires $[24,72]$, where calibration Macro-F1 decreases from $0.169$ at $[12,36]$ to $0.016$.
\fi
The calibration shows dataset-dependent knees: BBC and Udemy select $[6,18]$ Early Hints, Amazon and Reddit $[12,36]$, and Wikipedia $[24,72]$; the selected configurations yield calibration Macro-F1 $\leq 0.121$ across all five datasets.
All selected configurations use one PING, with HPACK randomization enabled for Amazon, Reddit, and Wikipedia.

\begin{table}[t!]
\caption{Server-Side Defense Landscape with \httptwo. Comparison of deploying the strongest defense on the leakiest server compared to H2PS $1^\text{st}$ Party (\benchkey1-4). Macro-F1 and Top-5 are means; their 95\% CIs are below 0.01 and omitted.}
\label{tab:server_mod_realworld_anon_sets}
\centering
    \begin{tabular}{@{}ccrrr|rr@{}}
    \toprule
    \textbf{Dataset}        & \textbf{Defense}      & \multicolumn{1}{c}{\textbf{F1}}  & \multicolumn{1}{c}{\textbf{Top-5}} & \multicolumn{1}{c}{\textbf{$\boldsymbol{\mathcal{K}^{*}}$}} &  \multicolumn{1}{c}{$\mathbf{\Delta \text{Down}}$} &  \multicolumn{1}{c}{$\mathbf{\Delta T}$} \\ \midrule
    \multirow{2}{*}{Amz.}
                             & TAM (CDN)      & $0.26$ & $0.51$ & $24.5$ & $2.2$ & $1.5$ \\
                             & H2PS           & $0.41$ & $0.73$ & $29.2$ & $2.4$ & $1.1$ \\
    \midrule
    \multirow{2}{*}{BBC}
                             & ALP (CDN)      & $0.87$ & $0.99$ & $2.3$ & $12.1$ & $0.5$ \\
                             & H2PS           & $0.18$ & $0.39$ & $42.5$ & $2.4$ & $0.0$ \\
    \midrule
    \multirow{2}{*}{Reddit}
                             & ALP (1st)      & $0.33$ & $0.45$ & $39.3$ & $3.4$ & $0.5$ \\
                             & H2PS           & $0.31$ & $0.49$ & $44.2$ & $1.2$ & $0.8$ \\
    \midrule
    \multirow{2}{*}{Udemy}
                             & TAM (1st)      & $0.41$ & $0.54$ & $14.6$ & $3.7$ & $1.5$ \\
                             & H2PS           & $0.52$ & $0.79$ & $13.2$ & $1.5$ & $0.1$ \\
    \midrule
    \multirow{2}{*}{Wiki}
                             & ALP (1st)      & $0.09$ & $0.17$ & $76.3$ & $5.0$ & $1.4$ \\
                             & H2PS           & $0.05$ & $0.15$ & $94.7$ & $2.3$ & $2.5$ \\
\bottomrule
    \end{tabular}

\end{table}

\Cref{tab:server_mod_realworld_anon_sets} summarizes the privacy--overhead trade-off of H2PS against the strongest single-server baseline (\textbf{\benchkey1--4}).
For each dataset, the baseline defense (ALPaCA or SRV-Tamaraw) is deployed on the leakiest server, which may be either the $1^\text{st}$-party server or a CDN; H2PS, in contrast, is always deployed only at the $1^\text{st}$-party
server and requires no third-party cooperation.
Despite this constraint, H2PS keeps the attacker's estimated candidate set above $13$ webpages across all five datasets.
Its protection is also more consistent across datasets: H2PS reaches $\mathcal{K}^{*}=13.2$--$94.7$, whereas the corresponding leakiest-server baselines range from $2.3$ to $76.3$.
The gains are most pronounced on BBC and Wikipedia, while H2PS remains competitive on Reddit.
Amazon presents a mixed trade-off: directly defending the leaky CDN yields lower F1 and Top-5, while H2PS achieves a larger candidate set ($29.2$ vs.\ $24.5$).
On Udemy, the leakiest-server baseline retains a small privacy advantage.

H2PS also offers a favorable overhead profile.
On four of five datasets, it reduces downstream overhead by $54$--$80\%$ relative to the corresponding leakiest-server baseline; on Amazon, the two are comparable ($2.4$ vs.\ $2.2$).
Latency is lower on Amazon, BBC, and Udemy, while Reddit and Wikipedia trade additional latency for comparable or stronger privacy.
Overall, H2PS provides more consistent protection using only the $1^\text{st}$-party server, without requiring cooperation from third-party or CDN servers.

%% file: sections/6_conclusion.tex
\section{Conclusion}
\label{section:conclusion}

We show that \httptwo supports practical subpage fingerprinting defenses without requiring a multi-hop anonymity network.
We demonstrate how established client- and server-side defenses can be emulated using \httptwo primitives and identify additional protocol features that improve their privacy--overhead trade-offs and deployment coverage.
Further, we highlight overlooked \httptwo features that can significantly reduce client-side defense overhead or expand the coverage of server-side defenses.

\Cref{tab:defenses_takeaway} summarizes these trade-offs.
On the \emph{client side}, CL-TAM achieves the strongest privacy in several datasets but with high overhead and substantial cross-dataset variability, whereas H2PC provides a higher minimum candidate set ($7.31$) at lower cost and with page-wide coverage.
On the \emph{server side}, SRV-TAM and ALPaCA provide strong protection when deployed on the appropriate server, but remain placement-dependent and costly.
H2PS instead extends protection page-wide from the $1^\text{st}$-party server alone, maintaining a candidate set above $13$ webpages at lower overhead and without requiring coordination with third-party servers.

Finally, we introduced a blueprint for benchmarking and quality assurance, providing a structured way to assess fingerprinting defenses and their privacy--overhead trade-offs.
Our evaluation shows that defense calibration and attacker hyperparameter tuning are integral to this assessment.
More broadly, the variation in selected defense parameters and strongest attackers across datasets highlights the benefits of multi-dataset evaluation and suggests that defense configurations should be tailored to the target website, as conclusions drawn from other websites may not transfer reliably.

\begin{table}[t!]
\caption{Summary of the \httptwo WF defenses. For server-side baselines, $\mathcal{K}^{*}$ ranges use the best single-server placement ($1^\text{st}$ party or CDN) for each dataset.}
\label{tab:defenses_takeaway}
\centering
    \begin{tabular}{@{}cccc@{}}
    \toprule
    \textbf{Defense} & \textbf{$[\mathcal{K}^{*}_{\min}-\mathcal{K}^{*}_{\max}]$}   & \textbf{Overhead} & \textbf{Coverage} \\ \midrule
    
    \multicolumn{4}{c}{\textbf{Client-Side}} \\
    CL-TAM  & $[2.5 - 38.4]$  & High & \textbf{Page-wide} \\
    FRONT   & $[1.8 - 27.7]$  & High & \textbf{Page-wide} \\
    H2PC    & $[7.3 - 20.1]$  & Low  & \textbf{Page-wide} \\ \midrule
    
    \multicolumn{4}{c}{\textbf{Server-Side}} \\
    SRV-TAM & $[1.7 - 24.5]$  & High & Per-server \\
    ALPaCA  & $[2.3 - 76.3]$  & High & Per-server \\
    H2PS $1^\text{st}$P & $[13.2 - 94.7]$ & Low & \textbf{Page-wide} \\
    \bottomrule
    \end{tabular}
\end{table}

\begin{acks}
We thank the anonymous reviewers for their constructive feedback, which helped improve the paper.
\end{acks}

%% file: sections/7_appendix_A_openscience.tex
\appendix 


\section{Open Science} 
\label{appendix:open_science}

\noindent\textbf{Code availability.}
We release the code for defense calibration, the WF defenses auditing tool (\texttt{wfaudit}), real-world data collection and replay, the \httptwo client- and server-side defenses, and the dataset creation and benchmarking pipeline at
\url{https://github.com/bcebere/Understanding-the-Privacy-Preserving-Potential-of-HTTP2-Against-Webpage-Fingerprinting}.

%% file: sections/7_appendix_B_ethics.tex
\section{Ethical Considerations}
\ifextver
\noindent\textbf{Data Collection.}
All data used in this work was collected from publicly accessible websites using automated browsing without bypassing authentication, paywalls, or access controls.
No personal data, user accounts, or sensitive identifiers were collected; only client–server traffic from scripted, non-authenticated sessions was recorded.

For each case study, we saved the request order and the downloaded resources to replay the content locally under various client or server defenses.

\noindent\textbf{Stakeholders and Impacts.}
This work involves three primary stakeholder groups,  as follows:

\begin{itemize}[left=0pt,noitemsep]
    \item\emph{End Users.} Common end users, including privacy-sensitive populations (such as journalists, activists, and users of privacy-enhancing technologies), are the primary stakeholders.
    Our experiments did not involve real users or user-generated traffic.
    While publication of fingerprinting techniques may increase the risk of traffic analysis by adversaries, the defensive insights provided by this work aim to strengthen user privacy in the long term.

    \item\emph{\httptwo Clients or Servers Developers.} Developers of \httptwo libraries may be impacted by our findings.
    This work identifies potential privacy weaknesses and benefits when employing various \httptwo features.
    
    \item\emph{Researchers and Practitioners.}
    The research community benefits from reproducible measurements and benchmarks.
    To support responsible reuse, we release our code for data collection and the benchmarking framework (\Cref{appendix:open_science}).
    The authors bear responsibility for accurate threat modeling and harm mitigation.
\end{itemize}

\noindent\textbf{Research Impact.}
This work has both positive and negative impacts, affecting the stakeholders.

\begin{itemize}[left=0pt,noitemsep]
    \item \emph{Positive Impacts.} First, we emulate concrete defenses for strengthening \httptwo clients and servers. Second, our open-source implementation enables practitioners to audit, reproduce, and extend our findings (\Cref{appendix:open_science}).
    \item \emph{Negative Impacts.}
    As with prior fingerprinting research, the techniques discussed could be misused for surveillance, targeted advertising, or aggressive monitoring.
    These risks reflect the dual-use nature of traffic analysis research.
    
\end{itemize}

\noindent\textbf{Mitigations.}
To mitigate potential harms, we implement and evaluate defenses alongside attacks.
Specifically, we implement defensive mechanisms at both the client and the server levels, using various \httptwo features.

\else
All data used in this work was collected through automated, non-authenticated browsing of publicly accessible websites without bypassing authentication, paywalls, or access controls. We collected no personal data, user accounts, or user-generated traffic; only traffic generated by our scripted client--server sessions was recorded. 
Downloaded resources and request order were saved to enable subsequent local replay under the evaluated defenses.

As with prior webpage-fingerprinting research, the techniques studied here are dual-use and could facilitate traffic analysis or surveillance.
We mitigate this risk by developing and evaluating client- and server-side defenses alongside the attacks and by releasing our reproducibility artifacts. 
\fi

\ifextver
\section{Generative AI Usage}

Large Language Models were used for limited editorial assistance, related-work search support, and code review/debugging. 
All generated text was reviewed and edited by the authors, all references were independently verified, and all research code was written, verified, and validated by the authors. 
No LLM was used to generate original research ideas or results forming the scientific contributions of this paper. 
The authors take full responsibility for the accuracy, originality, and integrity of the work.
\fi

%% file: sections/7_appendix_D_implementation.tex
\section{Data Collection Methodology}
\label{appendix:data_collection_methodology}

Given that we benchmark defenses at both the server and the client levels, we cannot use the real website deployments for experiments. 
Instead, we collect browser traces for each webpage and replay them using the \httptwo client and server described in \Cref{appendix:client_server_details}.

\subsection{Baseline Browser Crawlers}
\label{appendix:browser_crawlers}
The browser traces are collected using a headless Chromium browser automated via Playwright. 
For each URL, a new browser context is created with caching disabled, and the page is loaded until the \texttt{domcontentloaded} event fires.
Two traces are recorded in parallel: a client-side trace logging each request's URL and headers, and a server-side trace logging each response's URL, status code, content type, response headers, and round-trip duration (measured as wall-clock time between request dispatch and response receipt). 
Both client and server browser traces are serialized to JSON and saved to disk.
The raw response body of each resource is saved to disk and referenced by path in the server trace JSON. 

These traces (request order and response content) are replayed in a Python \httptwo client - server environment, with various defenses enabled.
The code responsible for collecting the browser traces is available in ``datasets/browser\_crawlers'' in the code repository.

\subsection{ \httptwo Client - Server Simulation}
\label{appendix:client_server_details}

The client-server code used to replay the browser traces is available in the code repository, with a proof-of-concept standalone library `h2deflib`.
The \httptwo clients and server (and their defenses) are implemented using the \texttt{h2} Python library, version 4.1.0. 
The \texttt{h2} library is a pure-Python, fully compliant implementation (RFC 9113 \cite{DBLP:journals/rfc/rfc9113}) of the \httptwo protocol, that provides the low-level building blocks necessary to implement \httptwo clients and servers without requiring a specific I/O framework.

\subsection{PCAP Traces Capture}
\label{appendix:pcap_captures}

To create the datasets for each benchmark, we replay the previously captured browser traces using the client--server setup described in \Cref{appendix:client_server_details}. The client and server run in separate Docker containers connected through a Docker virtual network; traffic is therefore exchanged between the two container network namespaces rather than over the host loopback interface.

Concretely, the \texttt{client\_runner.py} script automates the capture of network traffic during replay. At startup, it loads the pre-captured browser data (\Cref{appendix:browser_crawlers}), including the requests and response bodies associated with each webpage. For each (webpage, repeat) pair, the script starts a Scapy \texttt{AsyncSniffer} on the client-side network interface and replays the webpage's \httptwo requests against the server container. 
Requests are grouped by connection (i.e., by domain), and the configured client-side and server-side defenses are applied during replay. Once the replay completes, the sniffer is stopped, and the captured packets are stored in a PCAP file for subsequent processing by the PCAP parsing pipeline.

\subsection{Evaluation Datasets Creation}
\label{appendix:evaluation_dataset_creation}

The PCAP traces are parsed and converted into two distinct evaluation datasets: a 2D representation and a 3D representation.

For 2D datasets (used by k-FP and WeFDE), each trace is loaded as a two-column CSV of timestamps and signed packet sizes, from which a flat 1D feature vector is extracted per trace covering packet counts, inter-packet timing, burst statistics, and CUMUL features. Stacking these vectors across all traces yields a 2D feature matrix of shape $(\text{n\_traces},\ \text{LIM}_{\text{conns}} * \text{n\_features})$. 

For 3D datasets (used by DF, VarCNN, RobustFP, Holmes and DeepSE-WF), each trace is converted into two channels: a signed timing channel ($\text{sign} \times \text{timestamp}$) and a signed size channel ($\text{sign} \times |\text{size}| / 2000$), each padded or truncated to \texttt{feature\_length} (set adaptively as the median non-zero trace length plus 50, capped at the specified maximum $5,000$). The two channels are stacked along axis 0 to give a per-trace tensor of shape $(2, \allowbreak \ \texttt{feature\_length})$. Collecting all traces yields a 3D tensor of shape $(\text{n\_samples},\allowbreak \ 2, \allowbreak \ \texttt{feature\_length})$, which is then standardized per channel across samples and time positions using the global mean and standard deviation.

The code responsible for parsing the traces and creating the datasets is available in the code repository, in ``wfaudit/src/wfaudit/parser.py''.

\section{Security Estimators Details}
\label{appendix:security_estimators_extra}
We employ two categories of security estimators: (1) \emph{fingerprinting classifiers} that measure defense effectiveness through classification performance, and (2) \emph{information-theoretic estimators} that quantify residual information leakage independently of specific attack strategies.

\subsection{The Machine-Learning Estimators}
We first describe the fingerprinting classifiers used in this study, together with their hyperparameter search spaces.
To account for distribution shifts introduced by the defenses, we tune classifier hyperparameters independently for each dataset and defense configuration.

\subsubsection{K-Fingerprinting~\cite{DBLP:conf/uss/HayesD16}} k-FP extracts hand-crafted features from labeled network traces and trains a random decision forest classifier. Each trace is then represented by the vector of leaf-node identifiers. This compact fingerprint is then matched against known fingerprints using a nearest-neighbor approach to identify the visited site. 

For hyperparameter search, we tune the number of trees in the random forest and the number of nearest neighbors used for classification. 
We search $n_{\mathrm{trees}}\in\{50, \allowbreak 100, \allowbreak \ldots, \allowbreak 500\}$ and $k\in\{2, \allowbreak \ldots, \allowbreak 15\}$. 
These parameters control, respectively, the complexity of the learned fingerprint representation and the resolution of the neighbor-based voting stage.

\subsubsection{Deep-FP~\cite{DBLP:conf/ccs/SirinamIJW18}} DF uses a deep convolutional neural network to learn features directly from raw traces, without relying on hand-crafted features (unlike K-FP). The original implementation is available at  \href{https://github.com/deep-fingerprinting/df}{github.com/deep-fingerprinting/df}.

The DF neural network architecture consists of four convolutional blocks, each with two \texttt{Conv1d} layers of kernel size 5 and same padding, ELU activations ($\alpha = 1.0$), and a dropout rate of $0.1$. The channel depths double across blocks: $32 \to 64 \to 128 \to 256$. A global average pooling layer is followed by a two-layer embedding head that projects to \texttt{embedding\_size}~$= 512$ dimensions via a \texttt{ReLU} activation and dropout ($p = 0.1$), and a classification head with dropout ($p = 0.5$).

For hyperparameter search, we tune the learning rate, weight decay, classifier dropout, and batch size. 
The learning rate is searched logarithmically over $[10^{-4},5\times10^{-3}]$; weight decay over $\{0,10^{-6},\allowbreak 10^{-5}, \allowbreak 10^{-4}, \allowbreak 10^{-3}, \allowbreak 10^{-2}\}$; classifier dropout over $[0.3,0.7]$; and batch size over $\{64, \allowbreak 128, \allowbreak 200, \allowbreak 256\}$.

\subsubsection{VarCNN~\cite{DBLP:journals/popets/BhatLKD19}} VarCNN is a data-efficient website-fingerprinting attack that uses a ResNet-based CNN, which leads to better performance with fewer training traces (compared to DF). The original implementation is available at \href{https://github.com/sanjit-bhat/Var-CNN}{github.com/sanjit-bhat/Var-CNN}.

The VarCNN classifier is built on a ResNet-18 backbone with four stages of $[2, 2, 2, 2]$ residual blocks, each consisting of two \texttt{Conv1d} layers of kernel size 3 with batch normalization ($\varepsilon = 10^{-5}$) and ReLU activations. The input embedding uses a \texttt{Conv1d} layer of kernel size 7 and stride 2, followed by batch normalization, ReLU, and max pooling (kernel size 3, stride 2). Channel depths double across stages: $64 \to 128 \to 256 \to 512$. Global average pooling is applied after the final stage, followed by a two-layer embedding head that projects $512 \to 1024 \to \texttt{embedding\_size}$ ($= 512$) with ReLU and dropout ($p = 0.1$), and a classification head with ReLU, dropout ($p = 0.1$), and a linear output layer. 

For hyperparameter search, we tune the learning rate, weight decay, dropout, and batch size.
The learning rate is searched logarithmically over $[10^{-4},5\times10^{-3}]$; weight decay over $\{0,10^{-6}, \allowbreak 10^{-5}, \allowbreak 10^{-4}, \allowbreak 10^{-3}, \allowbreak 10^{-2}\}$; dropout over $[0.1,0.5]$; and batch size over $\{64,128,200,256\}$.

\subsubsection{Holmes~\cite{DBLP:conf/ccs/Deng0024}} Holmes is a method focusing on early-stage website fingerprinting. The original implementation is available at \href{https://github.com/Xinhao-Deng/Website-Fingerprinting-Library}{github.com/Xinhao-Deng/Website-Fingerprinting-Library}.

The Holmes classifier uses a four-stage convolutional encoder (\texttt{conv\_num\_layers}~$= 4$), where each stage consists of a residual \texttt{ConvBlock1d} with two \texttt{Conv1d} layers of kernel size 3, same padding, batch normalization, and ReLU activations, plus a $1{\times}1$ projection shortcut when channel dimensions change. Between stages, max pooling (kernel size 3) and dropout ($p = 0.3$) are applied. Channel depths follow a doubling schedule capped at the embedding size: $128 \to 128 \to 128 \to 128$ (since \texttt{emb\_size}~$= 128$). Global average pooling is applied after the final stage, followed by a classification head with dropout ($p = 0.3$) and a linear output layer. 

For hyperparameter search, we tune the learning rate, weight decay, dropout, and batch size using the same ranges as VarCNN: learning rate in $[10^{-4},5\times10^{-3}]$ on a logarithmic scale, weight decay in $\{0, \allowbreak 10^{-6}, \allowbreak 10^{-5}, \allowbreak 10^{-4}, \allowbreak 10^{-3}, \allowbreak 10^{-2}\}$, dropout in $[0.1,0.5]$, and batch size in $\{64, \allowbreak 128, \allowbreak 200, \allowbreak 256\}$.

\subsubsection{RobustFP-CNN~\cite{DBLP:conf/uss/0001JG0Z023}} Robust Fingerprinting combines a dedicated traffic representation with a CNN classifier. 
In our evaluation, we use its CNN architecture with our packet representation rather than reproducing the complete Robust Fingerprinting preprocessing pipeline; we therefore refer to this classifier as \emph{RobustFP-CNN}. 
The original Robust Fingerprinting implementation is available at \href{https://github.com/robust-fingerprinting/RF}{github.com/robust-fingerprinting/RF}.

The RobustFP-CNN classifier processes input through two sequential stages. The first is a 2D convolutional frontend with two pairs of \texttt{Conv2d} layers (kernel size $(3, 6)$, same padding), channel depths $1 \to 32 \to 64$, max pooling and dropout ($p = 0.1$) between pairs. The second is a 1D convolutional backend with channel configuration $[128, \allowbreak 128, \allowbreak \texttt{M}, \allowbreak 256, \allowbreak 256, \allowbreak \texttt{M}, \allowbreak 512, \allowbreak n_\text{classes}]$, where \texttt{M} denotes max pooling with dropout ($p = 0.3$), and each layer uses kernel size 3. Global average pooling produces the final logits.

For hyperparameter search, we tune the learning rate, weight decay, convolutional dropout, and batch size. The learning rate is searched logarithmically over $[10^{-4},5\times10^{-3}]$; weight decay over $\{0, \allowbreak 10^{-6}, \allowbreak 10^{-5}, \allowbreak 10^{-4}, \allowbreak 10^{-3}, \allowbreak 10^{-2}\}$; convolutional dropout over $[0.1,0.5]$; and batch size over $\{64, \allowbreak 128, \allowbreak 200, \allowbreak 256\}$.

All CNN-based methods are modified to also process the packet-length information, which is informative in our datasets (and constant per \tor cell in the original implementations).


\subsection{\textbf{Information Leakage Estimators}}

\subsubsection{WeFDE~\cite{DBLP:conf/ccs/LiGH18}} estimates the mutual information using manually selected features. The original implementation is available at \href{https://github.com/s0irrlor7m/InfoLeakWebsiteFingerprint}{github.com/s0irrlor7m/InfoLeakWebsiteFingerprint}, and a Python version is available at \href{https://github.com/notem/reWeFDE}{github.com/notem/reWeFDE}. 

The WeFDE information leakage estimator uses the following hyperparameters: Bandwidth selection in the KDE uses the Hall plug-in method, with a rule-of-thumb fallback; Individual feature leakages are estimated using \texttt{n\_samples}~$= 5{,}000$ Monte Carlo samples, while the final joint cluster leakage uses \texttt{n\_samples}~$= 50{,}000$. The top \texttt{topn}~$= 20$ features by individual leakage are selected for joint analysis, after pruning redundant features whose normalized mutual information exceeds \texttt{nmi\_threshold}~$= 0.7$. 

\subsubsection{DeepSE-WF~\cite{DBLP:journals/popets/VeichtRB23}} estimates the mutual information and the  Bayes error by using specialized kNN-based estimators on learned latent feature spaces. The original implementation is available at \href{https://github.com/veichta/DeepSE-WF}{github.com/veichta/DeepSE-WF}. 

DeepSE-WF estimates mutual information by training an embedding model and applying $k$-NN estimators. The embedding backbone is DF with \texttt{embedding\_size}~$= 512$, \texttt{dropout}~$= 0.1$, trained with \texttt{batch\_size}~$= 200$ and input sequences of length~$= 5{,}000$. The dataset is split into train, validation, and two held-out test sets via stratified $k$-fold cross-validation (\texttt{k\_fold}~$= 5$); in each fold, the embedding model is trained on the training split and the two test splits are used to compute pairwise $k$-NN distance matrices. MI is then estimated in both directions (test1$\to$test2 and test2$\to$test1) and averaged. The $k$-NN estimator uses \texttt{squared\_l2} distance, with \texttt{mi\_k}~$= 5$ neighbours for mutual information. 

All these models are available in the ``wfaudit'' folder in the repository.

%% file: sections/7_appendix_E_calibration.tex
\clearpage
\section{Security Estimators Hyperparameter Tuning}
\label{appendix:attacker_calibration}

We implement hyperparameter optimization using Optuna.
We tune each attacker independently for each dataset--defense pair using a separate Optuna study, with Macro-F1 on a stratified held-out validation split as the optimization objective. To reduce tuning cost, each search uses up to 150 traces per webpage, of which 20\% are reserved for validation.

Before optimization, we evaluate the attacker's default configuration on the same tuning split. 
We retain the optimized parameters only when the best Optuna trial achieves a higher Macro-F1 than the default configuration; otherwise, we preserve the default parameters. 
We then retrain the selected configuration and evaluate it on the full dataset using the same benchmarking procedure as the remaining experiments.

The following tables (\Cref{tab:hp-amazon}, \Cref{tab:hp-bbc}, \Cref{tab:hp-reddit}, \Cref{tab:hp-udemy}, \Cref{tab:hp-wikipedia}) report, for each dataset--defense pair, the strongest attacker after tuning and the hyperparameters selected for that attacker.
We abbreviate batch size as \emph{bs}, dropout as \emph{do}, learning rate as \emph{lr}, and weight decay as \emph{wd}.

\begin{table}[t]
\centering
\caption{Hyperparameter tuning on the \emph{Amazon Dataset}. Abbreviations: batch size $\to$ bs; dropout $\to$ do; learning rate $\to$ lr; weight decay $\to$ wd.}
\label{tab:hp-amazon}
\small
\setlength{\tabcolsep}{4pt}
    \begin{tabular}{@{}l l l@{}}
    \toprule
    Defense & Best Model & Selected parameters \\
    \midrule
    \multicolumn{3}{@{}l}{\emph{Client-side}} \\
    HTTPOS             & RobustFP-CNN  & bs 64, do .448, lr 9.2e-4, wd 0 \\
    LLaMA              & Holmes        & bs 256, do .147, lr 4.6e-3, wd 1.0e-6 \\
    FRONT              & Holmes        & bs 256, do .147, lr 4.6e-3, wd 1.0e-6 \\
    CL-Tamaraw         & Holmes        & bs 256, do .147, lr 4.6e-3, wd 1.0e-6 \\
    H2PC               & Holmes        & bs 256, do .147, lr 4.6e-3, wd 1.0e-6 \\
    \addlinespace
    \multicolumn{3}{@{}l}{\emph{Server-side}} \\
    SRV-ALPaCA (1st)   & RobustFP-CNN  & bs 64, do .457, lr 8.6e-4, wd 0 \\
    SRV-ALPaCA (3rd)   & Holmes        & bs 256, do .147, lr 4.6e-3, wd 1.0e-6 \\
    SRV-ALPaCA (all)   & Holmes        & bs 256, do .147, lr 4.6e-3, wd 1.0e-6 \\
    SRV-Tamaraw (1st)  & RobustFP-CNN  & bs 64, do .165, lr 3.4e-4, wd 1.0e-4 \\
    SRV-Tamaraw (3rd)  & Holmes        & bs 256, do .147, lr 4.6e-3, wd 1.0e-6 \\
    SRV-Tamaraw (all)  & Holmes        & bs 128, do .124, lr 1.1e-3, wd 1.0e-6 \\
    \bottomrule
    \end{tabular}
\end{table}

\begin{table}[t]
\centering
\caption{Hyperparameter tuning on the \emph{BBC Dataset}. Abbreviations: batch size $\to$ bs; dropout $\to$ do; learning rate $\to$ lr; weight decay $\to$ wd.}
\label{tab:hp-bbc}
\small
\setlength{\tabcolsep}{4pt}
    \begin{tabular}{@{}l l l@{}}
    \toprule
    Defense & Best Model & Selected parameters \\
    \midrule
    \multicolumn{3}{@{}l}{\emph{Client-side}} \\
    HTTPOS             & k-FP          & trees 350, k 9 \\
    LLaMA              & RobustFP-CNN  & bs 64, do .448, lr 9.2e-4, wd 0 \\
    FRONT              & RobustFP-CNN  & bs 64, do .457, lr 8.6e-4, wd 0 \\
    CL-Tamaraw         & Holmes        & bs 256, do .147, lr 4.6e-3, wd 1.0e-6 \\
    H2PC               & RobustFP-CNN  & bs 64, do .457, lr 8.6e-4, wd 0 \\
    \addlinespace
    \multicolumn{3}{@{}l}{\emph{Server-side}} \\
    SRV-ALPaCA (1st)   & RobustFP-CNN  & bs 64, do .457, lr 8.6e-4, wd 0 \\
    SRV-ALPaCA (3rd)   & RobustFP-CNN  & bs 64, do .448, lr 9.2e-4, wd 0 \\
    SRV-ALPaCA (all)   & RobustFP-CNN  & bs 64, do .448, lr 9.2e-4, wd 0 \\
    SRV-Tamaraw (1st)  & Holmes        & bs 256, do .147, lr 4.6e-3, wd 1.0e-6 \\
    SRV-Tamaraw (3rd)  & Holmes        & bs 128, do .124, lr 1.1e-3, wd 1.0e-6 \\
    SRV-Tamaraw (all)  & Holmes        & bs 256, do .147, lr 4.6e-3, wd 1.0e-6 \\
    H2PS (1st)         & Holmes        & bs 256, do .147, lr 4.6e-3, wd 1.0e-6 \\
    \bottomrule
    \end{tabular}
\end{table}

\begin{table}[t]
\centering
\caption{Hyperparameter tuning on the \emph{Reddit Dataset}. Abbreviations:
batch size $\to$ bs; dropout $\to$ do; learning rate $\to$ lr; weight decay $\to$ wd.}
\label{tab:hp-reddit}
\small
\setlength{\tabcolsep}{4pt}
    \begin{tabular}{@{}l l l@{}}
    \toprule
    Defense & Best Model & Selected parameters \\
    \midrule
    \multicolumn{3}{@{}l}{\emph{Client-side}} \\
    HTTPOS             & RobustFP-CNN  & bs 64, do .457, lr 8.6e-4, wd 0 \\
    LLaMA              & Holmes        & bs 128, do .124, lr 1.1e-3, wd 1.0e-6 \\
    FRONT              & RobustFP-CNN  & bs 64, do .457, lr 8.6e-4, wd 0 \\
    CL-Tamaraw         & Holmes        & bs 256, do .147, lr 4.6e-3, wd 1.0e-6 \\
    H2PC               & RobustFP-CNN  & bs 64, do .457, lr 8.6e-4, wd 0 \\
    \addlinespace
    \multicolumn{3}{@{}l}{\emph{Server-side}} \\
    SRV-ALPaCA (1st)  & RobustFP-CNN  & bs 64, do .448, lr 9.2e-4, wd 0 \\
    SRV-ALPaCA (3rd)  & Holmes        & bs 128, do .124, lr 1.1e-3, wd 1e-6 \\
    SRV-ALPaCA (all)  & RobustFP-CNN  & defaults kept \\
    SRV-Tamaraw (1st) & Holmes        & bs 128, do .124, lr 1.1e-3, wd 1e-6 \\
    SRV-Tamaraw (3rd) & Holmes        & bs 128, do .124, lr 1.1e-3, wd 1e-6 \\
    SRV-Tamaraw (all) & Holmes        & bs 128, do .124, lr 1.1e-3, wd 1e-6 \\
    \bottomrule
    \end{tabular}
\end{table}

\begin{table}[t]
\centering
\caption{Hyperparameter tuning on the \emph{Udemy Dataset}. Abbreviations: batch size $\to$ bs; dropout $\to$ do; learning rate $\to$ lr; weight decay $\to$ wd.}
\label{tab:hp-udemy}
\small
\setlength{\tabcolsep}{4pt}
    \begin{tabular}{@{}l l l@{}}
    \toprule
    Defense & Best Model & Selected parameters \\
    \midrule
    \multicolumn{3}{@{}l}{\emph{Client-side}} \\
    HTTPOS             & Holmes        & bs 128, do .124, lr 1.1e-3, wd 1.0e-6 \\
    FRONT              & Holmes        & bs 128, do .124, lr 1.1e-3, wd 1.0e-6 \\
    LLaMA              & Holmes        & bs 256, do .147, lr 4.6e-3, wd 1.0e-6 \\
    CL-Tamaraw         & Holmes        & bs 256, do .147, lr 4.6e-3, wd 1.0e-6 \\
    H2PC               & RobustFP-CNN  & bs 64, do .448, lr 9.2e-4, wd 0 \\
    \addlinespace
    \multicolumn{3}{@{}l}{\emph{Server-side}} \\
    SRV-ALPaCA (1st)  & RobustFP-CNN  & defaults kept \\
    SRV-ALPaCA (3rd)  & RobustFP-CNN  & bs 64, do .448, lr 9.2e-4, wd 0 \\
    SRV-ALPaCA (all)  & Holmes        & bs 256, do .148, lr 4.6e-3, wd 1e-6 \\
    SRV-Tamaraw (1st) & Holmes        & bs 128, do .124, lr 1.1e-3, wd 1e-6 \\
    SRV-Tamaraw (3rd) & Holmes        & bs 128, do .124, lr 1.1e-3, wd 1e-6 \\
    SRV-Tamaraw (all) & k-FP          & $n_\text{trees}$ 350, $k$ 9 \\
    \bottomrule
    \end{tabular}
\end{table}

\begin{table}[t]
\centering
\caption{Hyperparameter tuning on the \emph{Wikipedia Dataset}. Abbreviations:
batch size $\to$ bs; dropout $\to$ do; learning rate $\to$ lr; weight decay $\to$ wd.}
\label{tab:hp-wikipedia}
\small
\setlength{\tabcolsep}{4pt}
    \begin{tabular}{@{}l l l@{}}
    \toprule
    Defense & Best Model & Selected parameters \\
    \midrule
    \multicolumn{3}{@{}l}{\emph{Client-side}} \\
    HTTPOS             & RobustFP-CNN  & bs 64, do .448, lr 9.2e-4, wd 0 \\
    LLaMA              & RobustFP-CNN  & bs 64, do .457, lr 8.6e-4, wd 0 \\
    FRONT              & Holmes        & bs 256, do .147, lr 4.6e-3, wd 1.0e-6 \\
    CL-Tamaraw         & Holmes        & bs 256, do .147, lr 4.6e-3, wd 1.0e-6 \\
    H2PC               & RobustFP-CNN  & bs 64, do .448, lr 9.2e-4, wd 0 \\
    \addlinespace
    \multicolumn{3}{@{}l}{\emph{Server-side}} \\
    SRV-ALPaCA (1st)   & k-FP          & trees 50, k 3 \\
    SRV-ALPaCA (3rd)   & RobustFP-CNN  & bs 64, do .457, lr 8.6e-4, wd 0 \\
    SRV-ALPaCA (all)   & RobustFP-CNN  & defaults kept \\
    SRV-Tamaraw (1st)  & RobustFP-CNN  & bs 64, do .457, lr 8.6e-4, wd 0 \\
    SRV-Tamaraw (3rd)  & RobustFP-CNN  & bs 64, do .448, lr 9.2e-4, wd 0 \\
    SRV-Tamaraw (all)  & RobustFP-CNN  & bs 64, do .448, lr 9.2e-4, wd 0 \\
    \bottomrule
    \end{tabular}
\end{table}

%% file: sections/7_appendix_C_extras.tex
\section{Full Paper Version}
The full version of this paper~\cite{XXX} includes detailed defense calibration analyses, additional implementation details for the \httptwo client--server setup and dataset collection, and attacker hyperparameter-tuning details.